\documentclass[12pt,preprint]{aastex}

\newcommand{\kms}{\mbox{$\>{\rm km\, s^{-1}}$}}

\newcommand{\pc}{\>{\rm pc}}
\newcommand{\kpc}{\mbox{$\>{\rm kpc}$}}

\newcommand{\msun}{\>{\rm M_{\odot}}}

\newcommand\degrees{^\circ}
\newcommand{\epot}{\mbox{$\epsilon_\Phi$}}
\newcommand{\mdisk}{\mbox{$M_{disk}$}}

\def\etal{{et al.}}
\def\eg{{\it e.g.}}

\def\ie{{\it i.e.}}
\def\cf{{\it cf.}}

\slugcomment{Draft to ApJ} 
\lefthead{Debattista \etal} 

\righthead{The Causes of Halo Shape Changes Induced by Cooling
Baryons}

\begin{document}   

\title{The Causes of Halo Shape Changes Induced by Cooling Baryons:\\
Disks Versus Substructures}

\author{Victor P. Debattista\altaffilmark{1,2}, Ben
Moore\altaffilmark{3}, Thomas Quinn\altaffilmark{1}, Stelios
Kazantzidis\altaffilmark{4}, Ryan Maas\altaffilmark{1}, Lucio
Mayer\altaffilmark{3}, Justin Read\altaffilmark{3}, Joachim
Stadel\altaffilmark{3}}

\altaffiltext{1}{Astronomy Department, University of Washington, Box
351580, Seattle, WA 98195, {debattis;trq;maasr@astro.washington.edu.}}

\altaffiltext{2}{Brooks Prize Fellow}

\altaffiltext{3}{Department of Theoretical Physics, University of
Z\"urich, Winterthurerstrasse 190, CH-8057, Z\"urich, Switzerland {
moore;lucio;justin;stadel@physik.unizh.ch.}}

\altaffiltext{4}{Kavli Institute for Particle Astrophysics and
Cosmology, Department of Physics, Stanford University, P.O. Box 20450,
MS 29, Stanford, CA 94309; stelios@slac.stanford.edu.}


\begin{abstract}
Cold dark matter cosmogony predicts triaxial dark matter halos,
whereas observations find quite round halos.  This is most likely due
to the condensation of baryons leading to rounder halos.  We examine
the halo phase space distribution basis for such shape changes.
Triaxial halos are supported by box orbits, which pass arbitrarily
close to the density center.  The decrease in triaxiality caused by
baryons is thought to be due to the scattering of these orbits.  We
test this hypothesis with simulations of disks grown inside triaxial
halos.  After the disks are grown we check whether the phase space
structure has changed by evaporating the disks and comparing the
initial and final states.  While the halos are substantially rounder
when the disk is at full mass, their final shape after the disk is
evaporated is not much different from the initial.  Likewise, the halo
becomes (more) radially anisotropic when the disk is grown, but the
final anisotropy is consistent with the initial.  Only if the baryons
are unreasonably compact or massive does the halo change irreversibly.
We show that the character of individual orbits is not generally
changed by the growing mass.  Thus the central condensation of baryons
does not destroy enough box orbits to cause the shape change.  Rather,
box orbits merely become rounder along with the global potential.
However, if angular momentum is transferred to the halo, either via
satellites or via bars, a large irreversible change in the halo
distribution occurs.  The ability of satellites to alter the phase
space distribution of the halo is of particular concern to galaxy
formation simulations since halo triaxiality can profoundly influence
the evolution of disks.
\end{abstract}

\keywords{galaxies: evolution --- galaxies: formation --- galaxies:
halos --- dark matter}

%
%

\section{Introduction}
\label{sec:intro}

The dark matter halos that form via hierarchical growth in the cold
dark matter (CDM) cosmologies are generally triaxial with mean axial
ratios $b/a \sim 0.6$ and $c/a \sim 0.4$, where $c < b < a$ are the
short, intermediate, and long axes, respectively \citep{bbks_86,
bar_efs_87, frenk_etal_88, dub_car_91, jin_sut_02, bai_ste_05,
all_etal_06}.
Observational constraints on halo shapes can be obtained from the
Milky Way \citep{iba_etal_01, joh_etal_05, helmi_04, fel_etal_06},
from polar ring galaxies \citep{sch_etal_83, sac_spa_90, iod_etal_03},
from X-ray isophotal shapes \citep{buo_can_94, buo_etal_02} \citep[but
see also ][]{die_sta_07}, and from gravitational lensing
\citep{kochan_95, bar_etal_95, koo_etal_98, ogu_etal_03}.  For disk
galaxies, or disks surrounding elliptical galaxies, the ellipticity of
the potential in the mid-plane, \epot, can be constrained through
photometry and/or kinematics of stars or gas
\citep[e.g.,][]{fra_dez_92, hui_van_92, kui_tre_94, fra_etal_94,
sch_etal_97, and_etal_01, debatt_03, bar_sel_03, wei_etal_08}.
The general consensus from these studies is that dark matter halos are
rounder than those predicted by collisionless CDM simulations.  But this
need not be in disagreement with CDM since the condensation of baryons
to the centers of halos has been shown to lead to rounder halos
\citep{dubins_94, kkzanm04}.  For example, \citet{kkzanm04} find that
the principal axis ratios increase by $\sim 0.2-0.4$ in the inner
regions (although triaxiality is not completely erased) extending to
almost the virial radius.

Slowly rotating triaxial structures can be supported by centrophilic
box orbits \citep{schwar_79, ger_bin_85, statle_87, udr_mar_94,
fri_mer_97, val_mer_98}.  Several studies have shown that when a black
hole is present scattering of box orbits is responsible for causing an
elliptical galaxy to become rounder, or at least axisymmetric
\citep{lak_nor_83, ger_bin_85, nor_etal_85, mer_qui_98, val_mer_98,
hol_etal_02, kal_etal_04}.  These scattering events lead to a large
number of orbits becoming chaotic.  Chaos by itself, however, need not
be a fundamental limit to forming long-lived triaxial structures:
using orbit superposition, \citet{poo_mer_02} were able to construct
long-lived triaxial models of nuclei even in the presence of a large
fraction $(\ga 50\%)$ of chaotic orbits.
If axisymmetrization does occur, \citet{ger_bin_85} predict that it is
largely confined to the center and occurs gradually.  The $N$-body
simulations of a cored system by \citet{mer_qui_98} instead found that
the axisymmetrization extends to the entire system and occurs on a
crossing time for black holes of mass $\sim 2\%$ of the galaxy's mass.
When instead the system is cuspy, \citet{hol_etal_02} found that black
holes do not lead to a global axisymmetrization of the system.
Triaxial structures in disks (\ie\ bars) can also be destroyed by
central mass concentrations (CMCs).  The main mechanism is again
scattering by the CMC.  Although the main bar-supporting orbit family,
the $x_1$ orbits \citep{contop_80}, is a centrophobic loop family,
stars librating about the closed $x_1$ orbits can still get close to
the center and then be scattered by a CMC.  Simulations have shown
that the required mass for a soft CMC (\ie\ one with a scale of a few
100 pc) is an unrealistically large $\sim 20\%$ of the disk mass,
while the mass required of a hard CMC (few parsecs or less scale) is
$\sim 5\%$ of the disk mass \citep{she_sel_04, ath_etal_05,
deb_etal_06}, which is much larger than typical supermassive black
holes.

Likewise, it has often been assumed that the loss of halo triaxiality
when baryons cool inside halos is partly or mostly due to the
destruction of box orbits, which pass arbitrarily close to the center
after a sufficiently long time.  The fate of box orbits in the
presence of disks is of interest for various reasons beside the shape
of the halo.  Box orbits play an important role in speeding up the
mergers of supermassive black holes at the centers of galaxies 
\citep{mer_poo_04}.  Moreover, box
orbits lead to radial anisotropy, whereas the destruction of box orbits
results in tangential anisotropy.  This in turn affects the
event rate and energies of dark matter detection experiments involving
both direct scattering and indirect annihilation from capture by the
Sun or the Earth \citep[see the review by][]{jun_etal_96}.

In order to help shed light on these issues, we test whether box orbit
scattering is responsible for triaxial halos becoming rounder when
baryons cool inside them.  We do this via simulations in which we first 
grow and then evaporate disks inside triaxial halos.  Such evaporation, while
obviously unphysical, allows us to directly assess the impact of disks
on halos by comparing the initial and final states.
After the disks are grown, we find that the halos become substantially
rounder and their kinematics radially anisotropic.  But comparing their
initial and final shapes when the disk mass is zero in both cases, we
find that the changes are largely reversible.  The destruction of box
orbits being irreversible, halos should not recover their initial
states if this is the main cause of the shape change.  We also show
that if angular momentum is transferred to the halo (via bars or
satellites), then the irreversible changes are substantially larger.

Section \ref{sec:numerics} of this paper discusses the $N$-body
methods used in this study.  Section \ref{sec:central} presents the
results of simulations with growing rigid central massive objects.  In
\S \ref{sec:angmomtran} we present simulations in which angular momentum
is transferred to the halo either by a live bar or by satellites.
Section \ref{sec:orbitanalysis} presents a preliminary analysis of the
orbital evolution for a subsample of the simulations.  Our conclusions
are presented in \S \ref{sec:conc}.


\section{Numerical Methods}
\label{sec:numerics}

The basis of this work is that box orbit destruction is an
irreversible process.  Rather than following all orbits, we
adiabatically grow and then evaporate a disk to show that the
distribution function of a halo is not substantially changed despite
the fact that the halo appears very different when the disk is at full
mass.  Of course, evaporating the disk is a purely numerical
contrivance, but this allows us to test for halo distribution function
changes directly.
Although classical mechanics are time-reversible, the random phases of
any scattered orbits ensure that simply evaporating the central mass
is not enough to return to the initial configuration.  This would only
be possible if we had a perfect integrator and if we had reversed all
velocities, which we did not do.

\begin{figure}
\centerline{
\includegraphics[angle=-90.,width=\hsize]{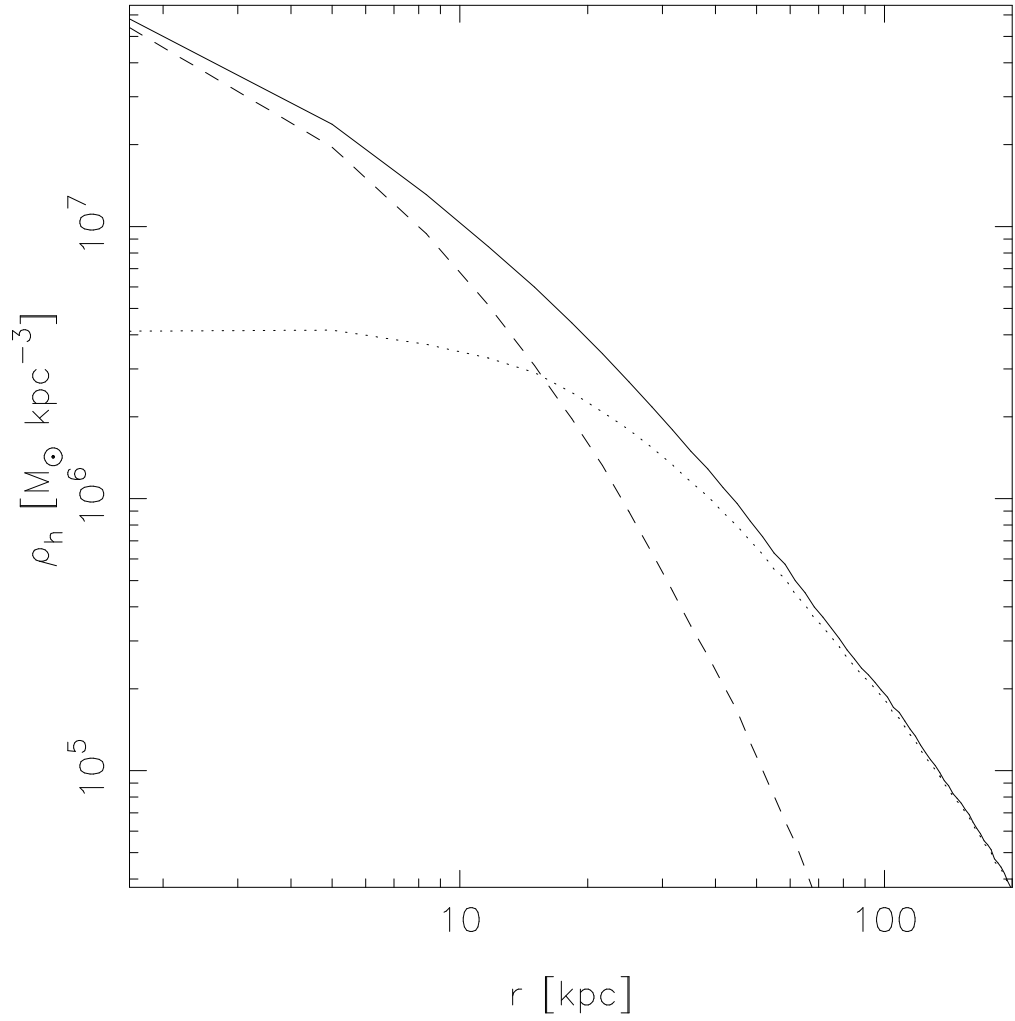}
}
\caption{Spherically averaged density profile of dark matter
particles in halo A before any baryons are introduced.  The solid line
is the full density profile, the dashed line is for the lower mass
species, while the dotted line is for the higher mass species.
\label{fig:segregation}}
\end{figure}

We formed prolate/triaxial halos via mergers, as described in
\citet{moo_etal_04}.  The initial spherical halos were generated from
a distribution function using the method described in \citet{kmm04}
with the added refinement that each halo is composed of two mass
species arranged on shells.  The outer shell has more massive
particles than the inner one, in order to increase the effective
resolution in the central parts.
Our model halo A was generated by the head-on merger of two prolate
halos, themselves the product of a binary merger of spherical systems.
The first merger placed the concentration $c=10$ halos 800 kpc apart
approaching each other at 50 \kms, while the second merger starts with
the remnant at rest, 400 kpc from an identical copy.  The resulting
halo is highly prolate with a mild triaxiality.  Halo model B was
produced by the merger of two spherical halos starting at rest, 800 kpc
apart.  Both halos A and B consist of $4\times 10^6$ particles.  The
outer particles are $\sim 18$ times more massive in halo A and $\sim
5$ times more massive in halo B.  A large part of the segregation by
particle mass persists after the mergers and the small radius regions
are dominated by low mass particles \citep[cf.][]{dehnen_05}.
Figure \ref{fig:segregation} shows the particle segregation in the
case of halo A.  We used a softening parameter $\epsilon = 0.1\kpc$
for all halo particles, although we have verified that using a larger
softening, $\epsilon = 1 \kpc$, for the more massive species does not
change our results.  Our force resolution was chosen to be smaller
than the vertical scale of the disk, thereby resolving short-range
forces.

Once we produced the prolate/triaxial halos, we inserted a disk of
particles that remains rigid throughout the experiments.  In all
mergers we have been careful to either give the halo no angular
momentum, or to place the disk's symmetry axis along the angular
momentum of the halo since otherwise additional evolution would result
\citep{deb_sel_99}.  The disks are composed of $300K$ equal-mass
particles each with a softening $\epsilon = 60-100\pc$.  The disk
distribution was, in all cases, exponential with scale length $R_{\rm
d}$ and Gaussian scale-height $z_{\rm d}/R_{\rm d} = 0.05$.  The disks
were placed at various orientations within the halos.  We refer to
these experiments by the halo axis along which the disk's symmetry
axis is aligned: in ``short-axis'' (S) experiments, the symmetry axis
of the disk is parallel to the short axis of the halo, while in
``long-axis'' (L) experiments, the symmetry axis of the disk is along
the halo's major axis.  If the halo is triaxial, then an
``intermediate-axis'' (I) experiment has the disk minor axis parallel
to the halo's intermediate axis.  Initially, the disk has negligible
mass, but this grows adiabatically linearly over time to a mass $M_b$
during a time $t_g$.  After this time, we slowly evaporated it during
a time $t_e$.  Thus,
\begin{equation}
\mdisk (t) = \left\{
\begin{array}{ll}
{M_b~ \frac{t}{t_g}} & {0 \leq t \leq t_g} \cr \cr 
{M_b~ (1 - \frac{t-t_g}{t_e})} & {t_g \leq t \leq t_g + t_e}.
\end{array}\right.
\end{equation}
From $t=0$ to $t_g+t_e$ the halo particles are free to move and
achieve equilibrium with the disk as its mass changes, but all disk
particles are frozen in place.  Since a triaxial global potential
leads to elliptical disks forming, we include one simulation with an
elliptical disk.

Another key assumption in these simulations is that the disks form
without much transfer of angular momentum to the halo.  While
formation of realistic galaxies requires that baryons conserve most of
their angular momentum \citep[e.g.,][]{som-lar_etal_99}, gas
condensation onto subhalos results in angular momentum transfer to the
halo \citep{nav_ste_97}.
We therefore present experiments in which a few softened
particles were introduced, with a mass grown in the same way.  We
refer to these experiments by the label ``P'' subscripted by ``f'' for
particles frozen in place and by ``l'' for live particles free to
move.  Lastly, we present one simulation, BA1, in which the disk at
$t_g$ is replaced by live particles and evolved for 10 Gyr before
evaporating the disk.  We set initial disk particle velocities for a
constant Toomre-$Q = 1.5$.  In setting up the disk kinematics we
azimuthally averaged radial and vertical forces.  Thus, our disk is
initially not in perfect equilibrium, but was close enough that it
quickly settled to a new equilibrium.

\begin{table}[!ht]
\begin{centering}
\begin{tabular}{cccccccc}\hline 
\multicolumn{1}{c}{Run} &
\multicolumn{1}{c}{Halo} &
\multicolumn{1}{c}{$r_{200}$} &
\multicolumn{1}{c}{$M_{200}$} &
\multicolumn{1}{c}{$M_b$} &
\multicolumn{1}{c}{$R_{\rm d}$} &
\multicolumn{1}{c}{$t_g$} &
\multicolumn{1}{c}{$t_e$} \\ 

 & & [kpc] & [$10^{12} \msun$] & [$10^{11} \msun$] & [kpc] & [Gyr] & [Gyr] \\ \hline

SA1 & A & 215 & 4.5 & 1.75 & 3.0 & 5 & 2.5 \\ 
SA2 & A & 215 & 4.5 & 5.25 & 3.0 & 5 & 2.5 \\ 
SA3 & A & 215 & 4.5 & 1.75 & 1.5 & 5 & 2.5 \\ 
IA1 & A & 215 & 4.5 & 1.75 & 3.0 & 5 & 2.5 \\ 
LA1 & A & 215 & 4.5 & 1.75 & 3.0 & 5 & 2.5 \\ 
LB1 & B & 106 & 0.65 & 1.05  & 3.0 &  15 & 7.0 \\ 
TA1 & A & 215 & 4.5 & 1.75 & 3.0 & 5 & 2.5 \\ 
EA1 & A & 215 & 4.5 & 1.75 & 3.0 & 5 & 2.5 \\ 
BA1 & A & 215 & 4.5 & 0.52 & 3.0 & 1.5 & 2.5 \\ 

P$_l$A1 & A & 215 & 4.5 & 1.75 & 0.5 & 5 & 2.5 \\ 
P$_l$A2 & A & 215 & 4.5 & 1.75 & 5.0 & 5 & 2.5 \\ 

P$_l$B1 & B & 106 & 0.65 & 0.7 & 3.0 & 10 & 4 \\ 
P$_f$B2 & B & 106 & 0.65 & 0.7 & 3.0 & 10 & 4 \\ 
P$_l$B3 & B & 106 & 0.65 & 0.35 & 0.1 &  5 & 5 \\ 

\hline
\end{tabular}
\caption{The simulations in this paper.  For the particle simulations
(P$_l$A1-P$_l$B3), $R_{\rm d}$ refers to the softening of the
particle(s).  For runs P$_l$A1 and P$_l$A2, the value of $M_b$ refers
to the combined mass of all the satellite particles at $t_g$.  The
disk in run TA1 is tilted by $30\degrees$ relative to the one in SA1
whereas the disk in EA1 is elliptical.  In run BA1, the disk at $t_g$
was replaced by live particles and evolved for 10 Gyr (during which time a
bar formed and then was destroyed), before being evaporated.}
\label{tab:simulations}
\end{centering}
\end{table}

\begin{figure}
\centerline{
\includegraphics[angle=-90.,width=0.5\hsize]{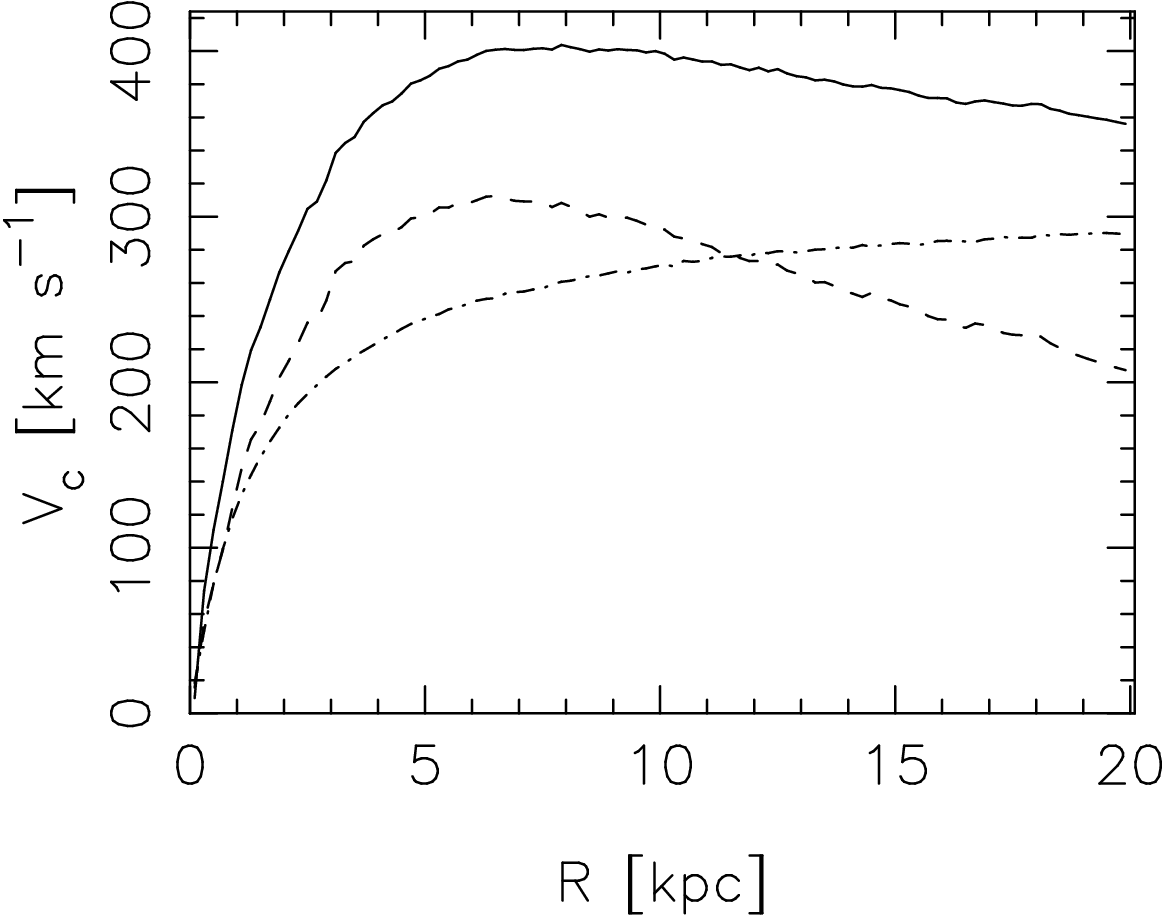}
\includegraphics[angle=-90.,width=0.5\hsize]{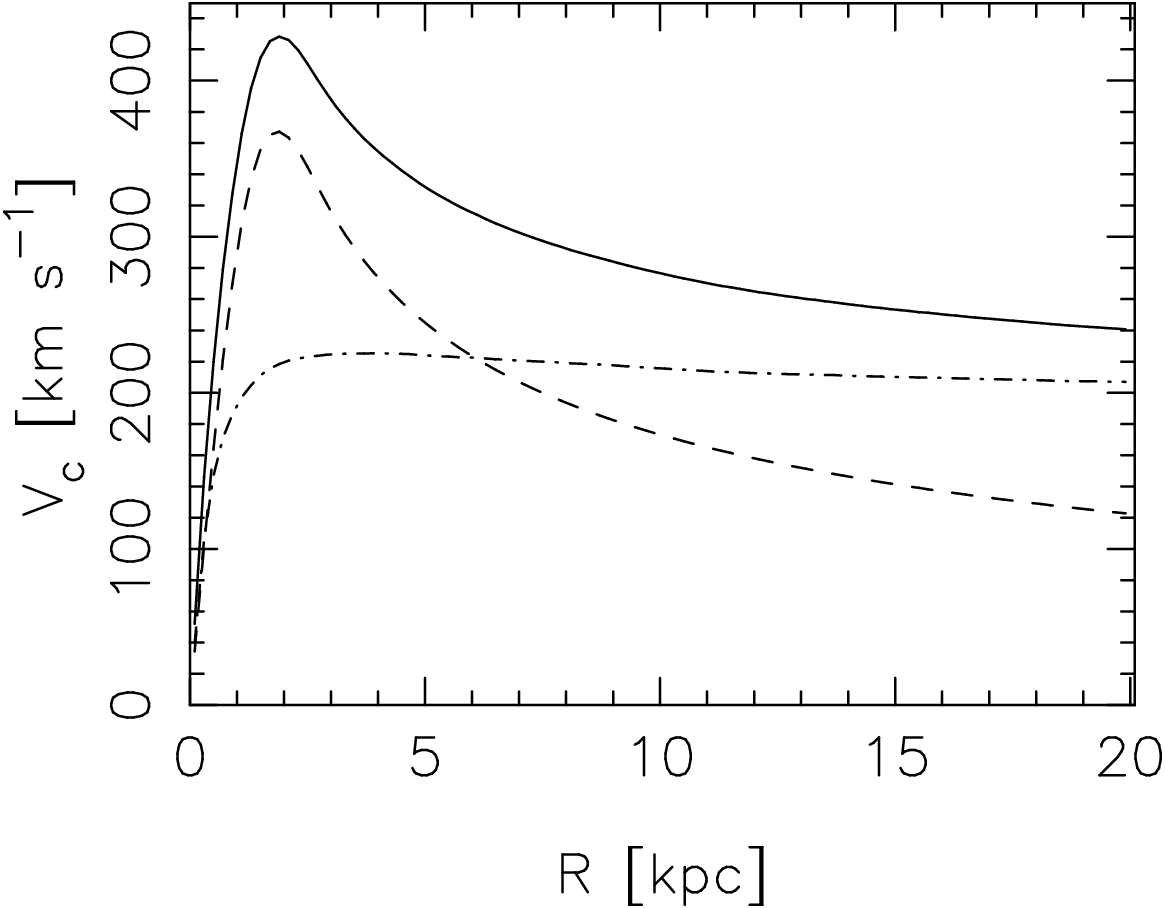}
}
\caption{Azimuthally averaged rotation curves of models SA1 ({\it left})
and P$_l$B1/P$_f$B2 ({\it right}) measured in the midplane.  In both panels
the solid line is the full rotation curve, the dashed line the
contribution of the baryons, and the dot-dashed line the contribution
from the halo.
\label{fig:rotcurves}}
\end{figure}

All the simulations in this paper, which are listed in Table 
\ref{tab:simulations}, were evolved with {\sc PKDGRAV}
\citep{stadel_phd}, an efficient, multi-stepping, parallel treecode.

\subsection{Measuring halo shapes}

To measure the axis ratios $c/a$ and $b/a$ we adopt a method based on
\citet{katz_91} that uses the eigenvalues of the (unweighted) moment
of inertia tensor $I$. For each bin of $N$ particles we computed
$I_{ij}$ as follows:
\begin{equation}
I_{ij} = \frac
{\sum_{k=1}^N m_{k} r_{i,k} r_{j,k}} 
{\sum_{k=1}^N m_k}.
\end{equation}
We then diagonalize $I$ and calculate 
\begin{equation}
b/a = \sqrt{{\cal I}_{22}/{\cal I}_{11}} \hspace{0.5cm} \mbox{and}\hspace{0.5cm} c/a
= \sqrt{{\cal I}_{33}/{\cal I}_{11}},
\end{equation}
where the ${\cal I}_{ii}$'s are the eigenvalues of $I$ and ${\cal
I}_{11} \geq {\cal I}_{22} \geq {\cal I}_{33}$.  A useful parameter
for expressing shape is the triaxiality parameter $T = (a^2 -
b^2)/(a^2 - c^2)$ \citep{fra_etal_91}.  The cases $T=0$ and $T=1$
correspond to oblate and prolate shapes, respectively, while $T=0.5$ is
the maximally triaxial case.

We measured shapes in shells of fixed semi-major axis widths around
the center of the system.  Thus, these shape measurements are
differential, rather than integrated \citep[\cf][]{katz_91}. We use
the iterative procedure of \citet{katz_91} in which the convergence
criterion is a variation in axis ratios by $<0.01\%$. In each
iteration the semi major axis of the shell is held fixed; a particle
is included in the calculation of $I_{ij}$ if $q_{lo} < q < q_{hi}$,
where $q$ is the ellipsoidal radius defined as
\begin{equation}
q ^{2} = x^{2} + \left ( \frac{y}{b/a} \right )^2 + \left (
\frac{z}{c/a} \right )^2.
\end{equation}
We used shell widths $q_{hi}-q_{lo} = 5 \kpc$ for all
models.

The center of the system is taken to be the center of mass of a sphere
of radius 1 \kpc\ centered on the minimum of the potential and is
fixed for all shells.  Tests performed in which the center of mass was
allowed to vary by up to 0.5 \kpc\ show less than 5\% variation in the
axis ratios past 10 \kpc.  Tests in which the limits of each shell
were reduced by half around the average radius of that shell gave axis
ratios that varied from the full resolution results by up to $0.08$
in the worst cases and by less than $0.05$ for most runs.  Shells were
not prevented from overlapping; as a result, some particles are sampled
in more than one shell.  We have verified that this does not bias our
shape estimates through the experiments with the halved shell widths,
where the shells never intersect.  When the number of particles in the
central shell is less than $10K$, then convergence is not reached after
20 iterations, or the axis ratios varies by as much as 20\%, so we
take this number to be a reasonable cutoff for the reliability of
these innermost shells and ignore shells with less particles. This
occurs in only two cases, and in general most inner shells have $>25K$
particles, which we find to be more than enough to ensure consistent
measurements with our method.


\section{Central Massive Objects}
\label{sec:central}
\thispagestyle{empty}
\setlength{\voffset}{-22mm}
\begin{figure*}
\centerline{
\includegraphics[width=0.45\hsize]{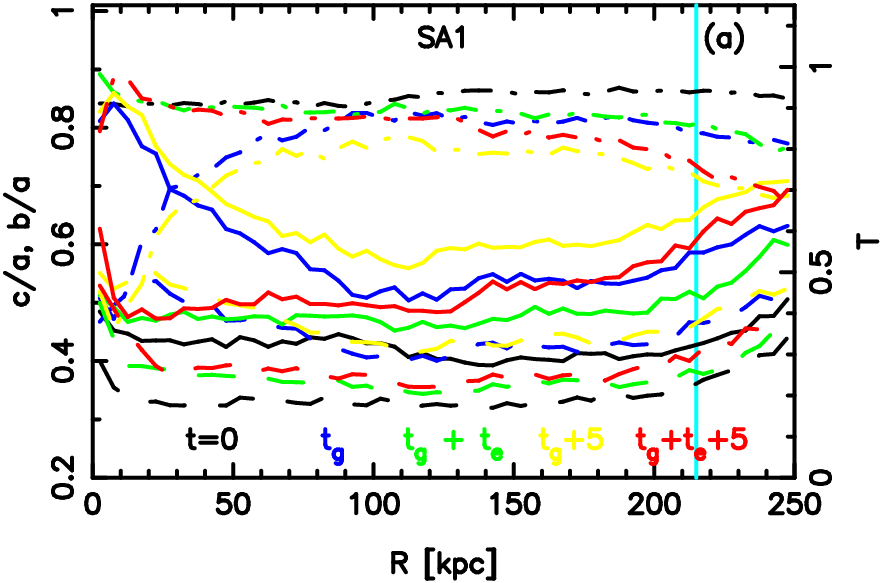}
\includegraphics[width=0.45\hsize]{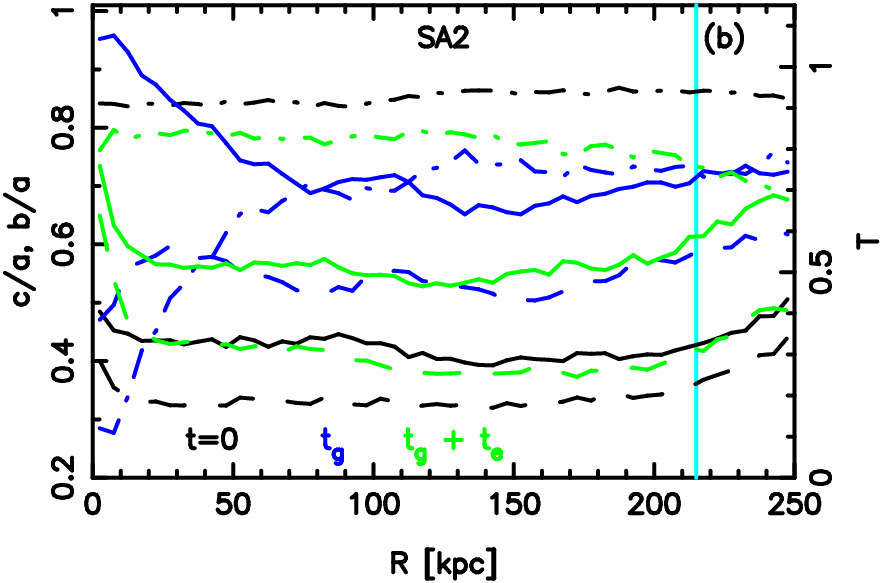}
}
\centerline{
\includegraphics[width=0.45\hsize]{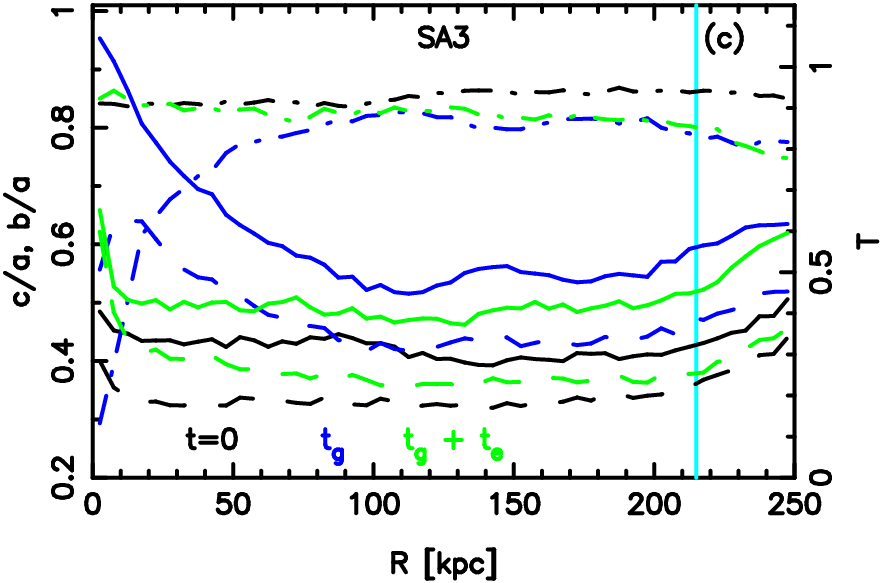}
\includegraphics[width=0.45\hsize]{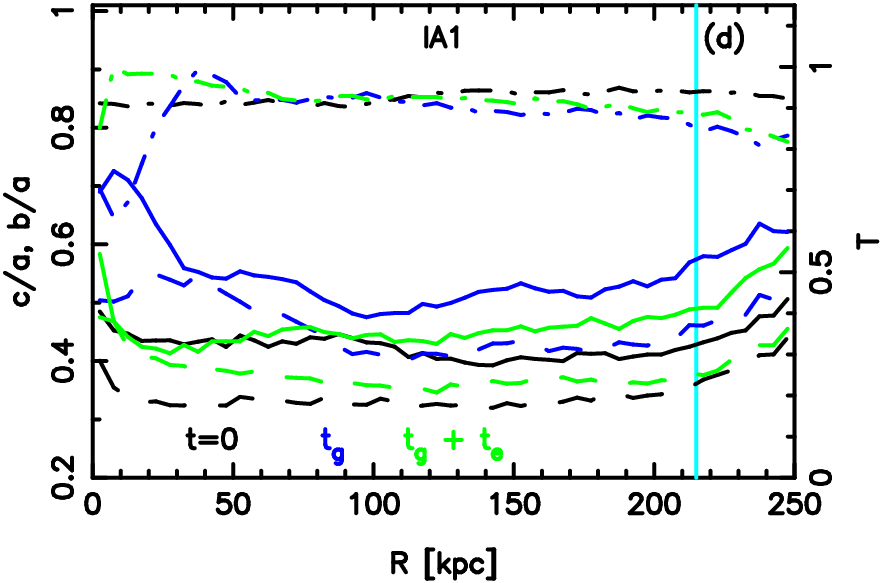}
}
\centerline{
\includegraphics[width=0.45\hsize]{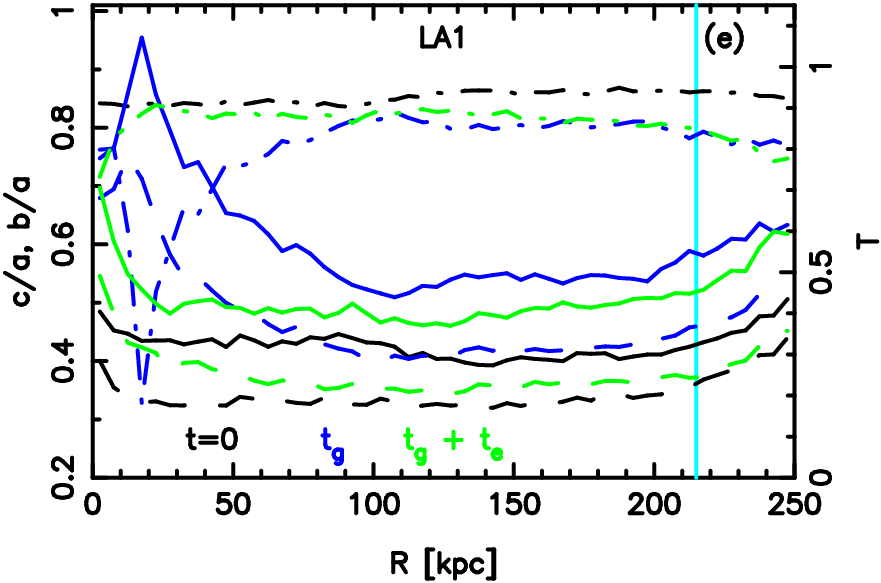}
\includegraphics[width=0.45\hsize]{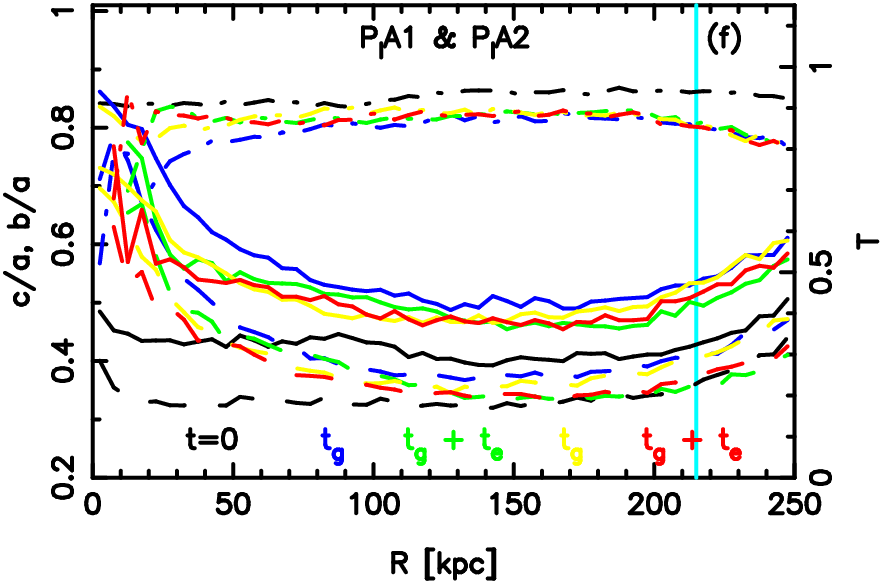}
}
\centerline{
\includegraphics[width=0.45\hsize]{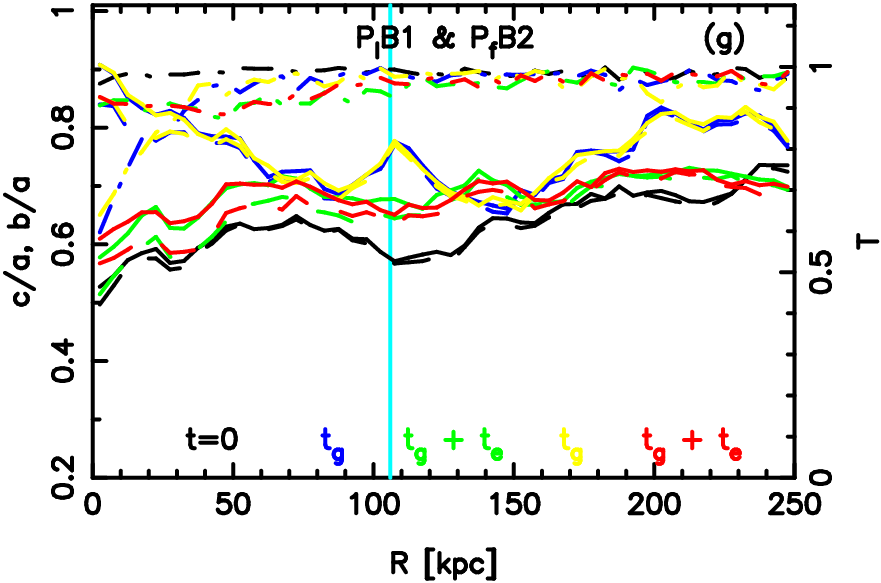}
\includegraphics[width=0.45\hsize]{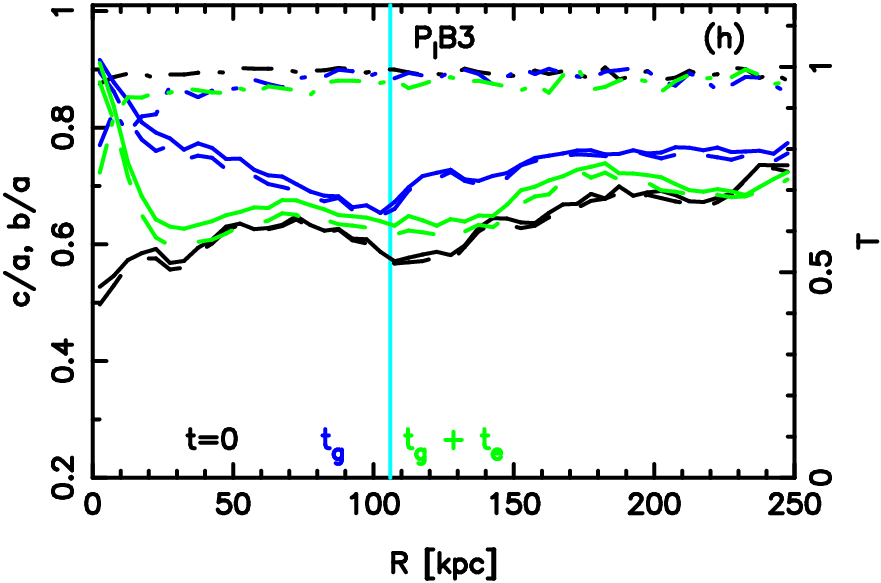}
}
\caption{Shape evolution in runs ({\it a}) SA1, ({\it b}) SA2, ({\it c}) SA3, ({\it d}) IA1,
({\it e}) LA1, ({\it f}) P$_l$A1 and P$_l$A2, ({\it g}) P$_l$B1 and P$_f$B2 and ({\it h})
P$_l$B3.  The solid lines show $b/a$, the dashed lines show $c/a$, and
the dot-dashed lines show $T$ (with scale indicated on the right-hand
side of each panel).  The black, blue, and green lines are at $t=0$,
$t_g$, and $t_g + t_e$, respectively.  In panel ({\it a}) black/yellow/red
shows the evolution if, after $t_g$, the disk is held at full mass for
a further 5 Gyr before it is evaporated.  In panel ({\it f}), the standard
colors are for P$_l$A1, while P$_l$A2 is indicated in black/yellow/red.
Likewise, in panel ({\it g}), the standard colors are for P$_f$B2, while
black/yellow/red are for P$_l$B1.  In all panels, the vertical cyan
line shows $r_{200}$.  The standard errors on the plotted axis ratios
(see text for details) are $<0.08$; shells with larger measurement
errors, generally at small radius, are not plotted.
\label{fig:shapes}}
\end{figure*}
\setlength{\voffset}{0mm}

\subsection{Short- and Intermediate-Axis Experiments}

\begin{figure}
\centerline{
\includegraphics[width=\hsize]{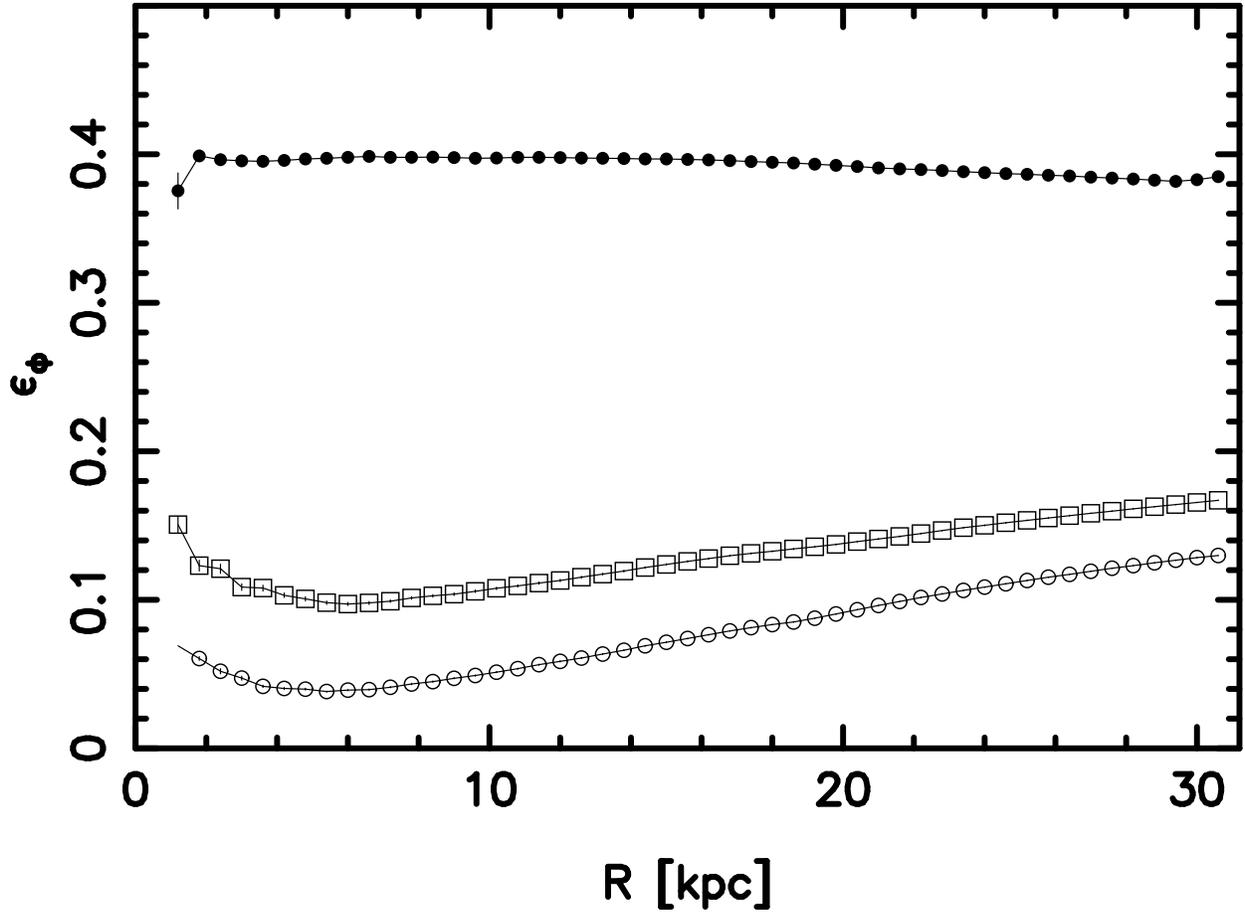}
}
\caption{Ellipticity of the potential, \epot, in the disk
mid-plane of run SA1.  The solid points are at $t=0$, and the open
points are at $t_g$, with squares for halo only and circles for
disk$+$halo.
\label{fig:potell}}
\end{figure}

\thispagestyle{empty}
\setlength{\voffset}{-22mm}
\begin{figure*}
\centerline{
\includegraphics[width=0.45\hsize]{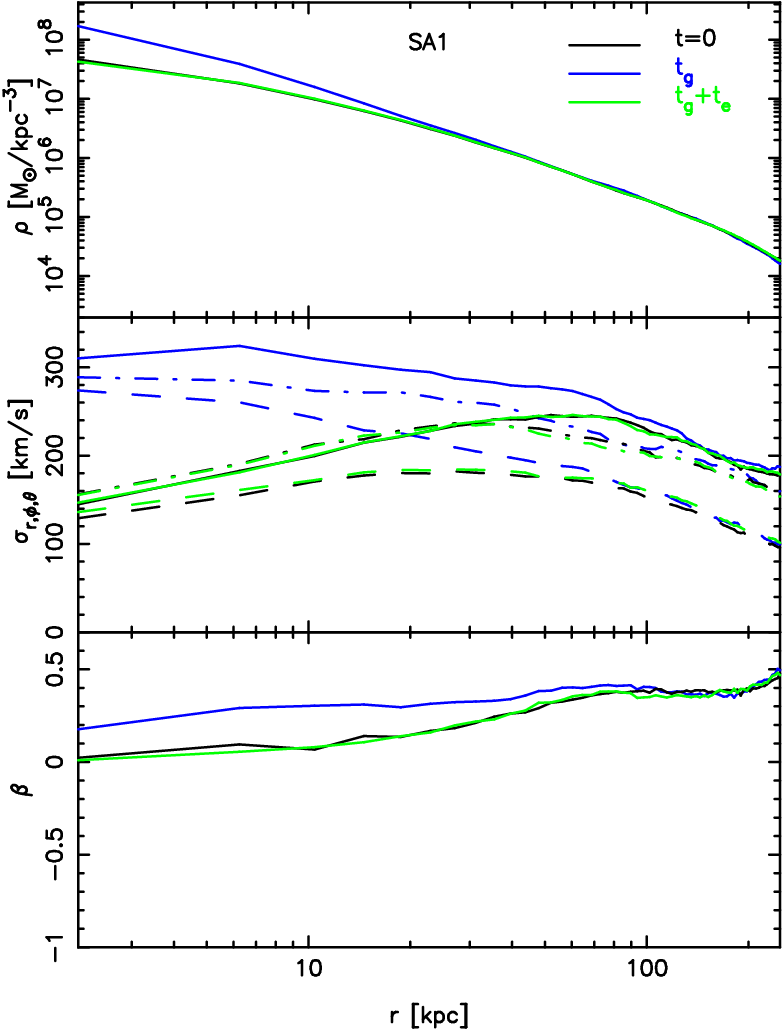}
\includegraphics[width=0.45\hsize]{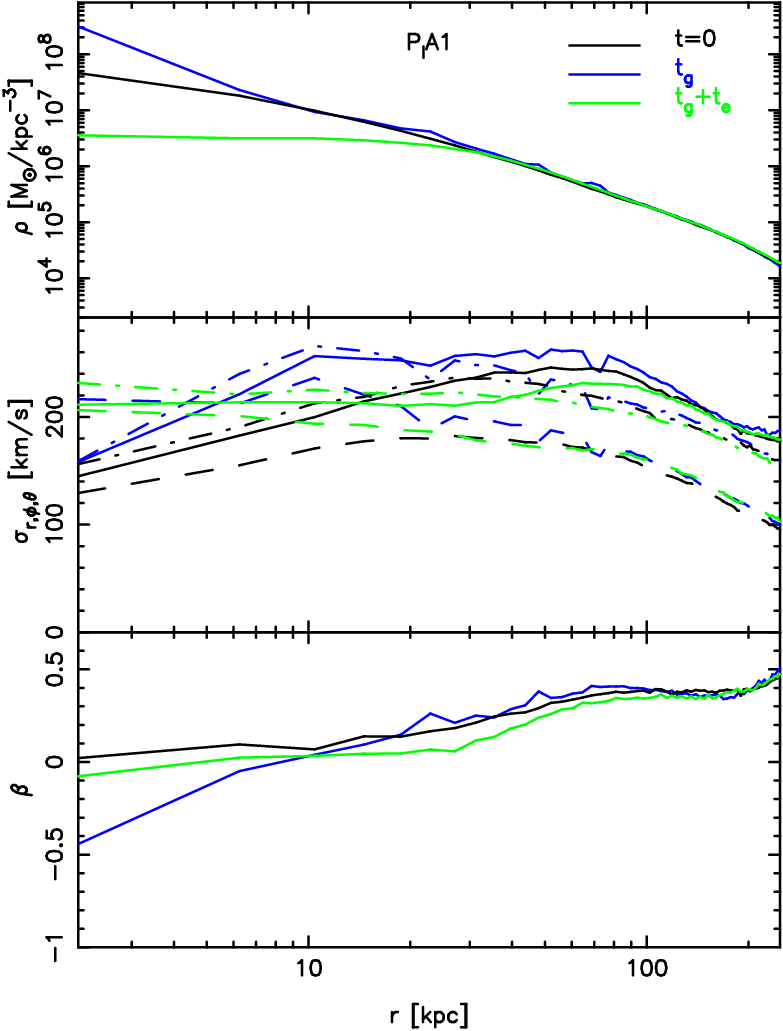}
}
\centerline{
\includegraphics[width=0.45\hsize]{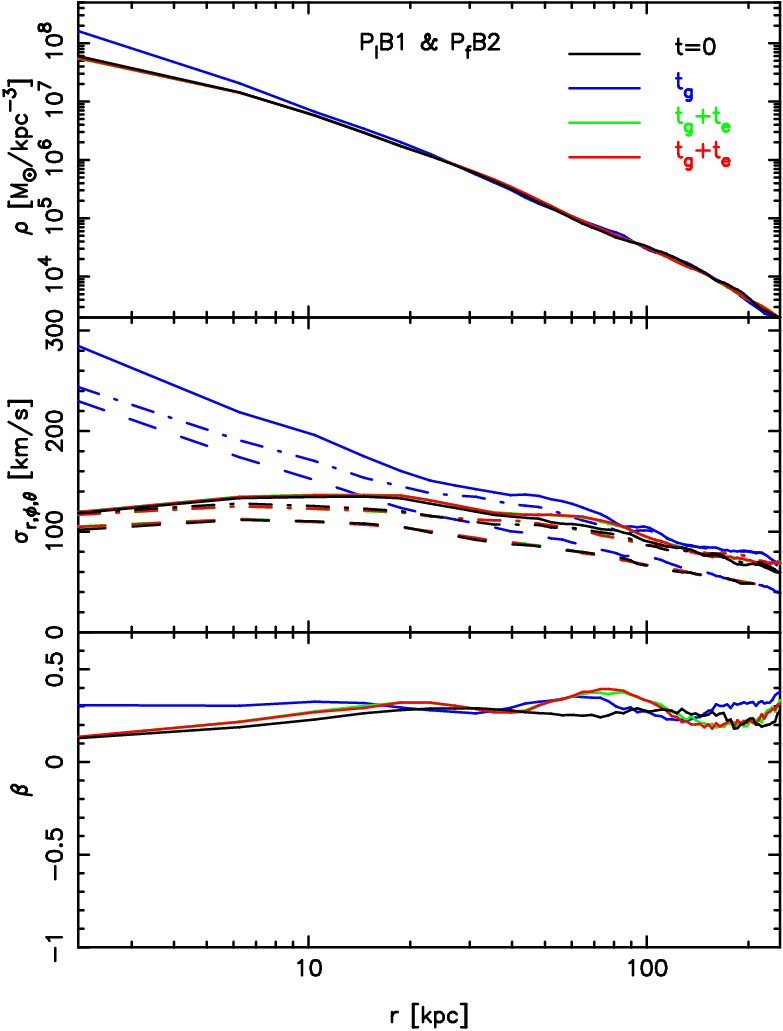}
\includegraphics[width=0.45\hsize]{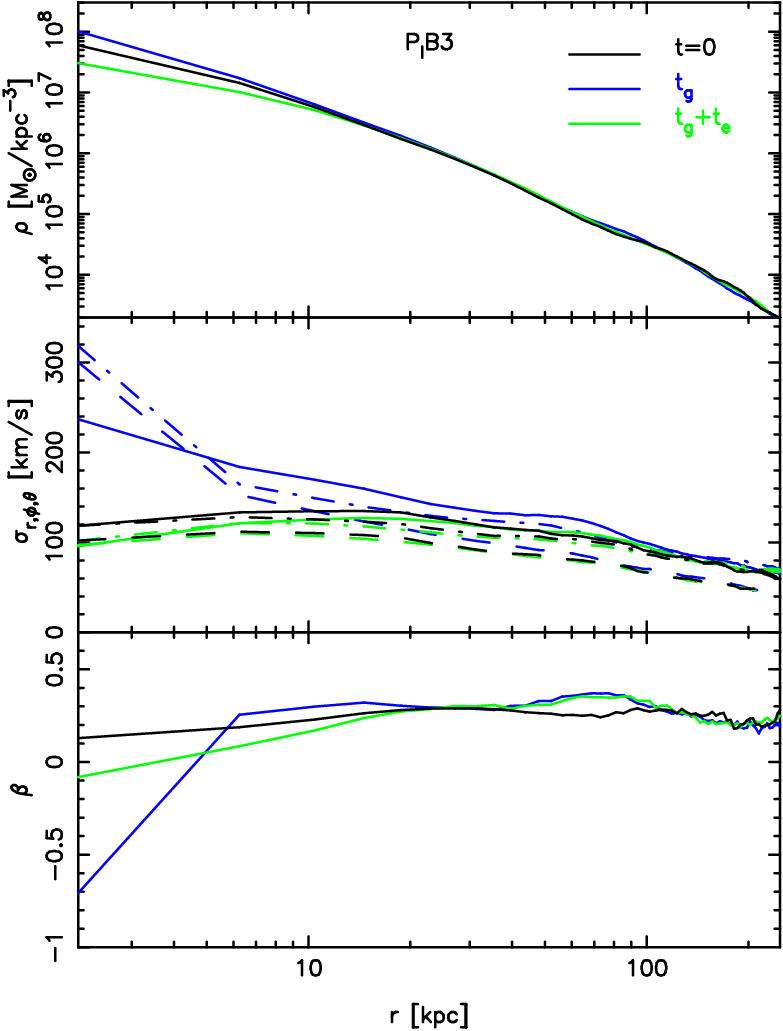}
}
\caption{Evolution of the spherically averaged density and
kinematics in runs SA1 ({\it top left}), P$_l$A1 ({\it top right}), P$_l$B1 and
P$_f$B2 ({\it bottom left}), and P$_l$B3 ({\it bottom right}).  The black, 
blue, and
green lines correspond to $t=0$, $t_g$, and $t_g+t_e$.  The red lines
show $t_g+t_e$ for P$_f$B2.  The top panels shows the densities.  In
the middle panels the solid lines indicate $\sigma_r$, the dashed
lines $\sigma_\phi$, and the dot-dashed lines $\sigma_\theta$.  Here
the $z$-axis from which the angle $\theta$ is measured is the long
axis of the halo.  The bottom panels show the anisotropy parameter,
$\beta$.
\label{fig:kinematics}}
\end{figure*}
\setlength{\voffset}{0mm}

In run SA1 we grew a disk inside halo A with the minor axes of the
disk and halo aligned.  This orientation is a natural one for disks to
form in since simulations have shown that the angular momenta of halos
are aligned with their {\it minor} axes \citep[\eg][]{war_etal_92,
por_etal_02, fal_etal_05}.  Once the disk is grown to its full mass it
dominates the inner rotation curve (see Fig
\ref{fig:rotcurves}, left).  The shape evolution of this highly prolate,
mildly triaxial halo is shown in Figure \ref{fig:shapes}a.  The effect
of the massive disk on the halo shape is large: having started out
with $b/a \la 0.45$ it becomes much rounder in the plane of the disk
($b/a \ga 0.6$ to 40 \kpc, \ie\ $\sim 0.2~ r_{200}$, the disk
constituting 17\% of the mass within this radius), as shown by the
blue lines.  The change in shape perpendicular to the disk is more
modest and the inner halo becomes significantly more triaxial than when it
started out.  In the disk plane, the combined potential starts out
very elliptical, $\epot \simeq 0.4$, and becomes quite round, with
$\epot < 0.1$ over the entire extent of the disk (Fig.
\ref{fig:potell}).  This $\epot$ is sufficiently small to be
consistent with the observed scatter in the Tully-Fisher relation
\citep{fra_dez_92}, even without the additional axisymmetrization of
the potential that would be caused by the disk's orthogonal response.
Once the disk is evaporated, the resulting halo shape, shown by the
green lines in Figure \ref{fig:shapes}a, is very similar to its
original shape, with the net increase in both $b/a$ and $c/a$ being
$\la 0.1$ throughout the inner 100 \kpc.  The final triaxiality is
barely changed from the starting one, despite the fact that the inner
halo was almost maximally triaxial at $t_g$.

Likewise, the final density and anisotropy profiles, shown in Figure
\ref{fig:kinematics}, are not significantly changed, despite the 
factor of $\sim 3.7$ increase in halo central density at $t_g$.  The 
halo anisotropy, $\beta = 1 - \sigma_t^2/\sigma_r^2$, where 
$\sigma_t^2 = \frac{1}{2}(\sigma_\theta^2 + \sigma_\phi^2)$, 
starts out $\beta \simeq 0$, grows to $\beta \ga 0.2$ at $t_g$ (\ie\
becomes radially anisotropic), and returns to $\beta \simeq 0$ at 
$t_g + t_e$.  If box orbits had been destroyed to any significant extent, 
we would have seen instead an increase in tangential anisotropy 
\citep[e.g.,][]{hol_etal_02}.

The small difference between $t=0$ and $t_g+t_e$ in halo shape and
kinematics suggests that the halo phase space distribution has not been
grossly altered by the presence of the massive disk.  There is little
evidence for a substantial amount of box orbit scattering, and any
chaos induced has to be quite mild.  All this is true despite the
quite large change in halo shape and kinematics when the disk is at 
full mass.

Figure \ref{fig:shapes}a ({\it yellow and red lines}) also shows the
evolution when we left the disk at full mass for 5 Gyr before
evaporating it.  The halo becomes slightly rounder (by about $b/a \la
0.1$) at all radii both after the additional 5 Gyr and once the disk
is evaporated.  The difference is largest inside $ \sim 30 \kpc$ where
the final $b/a$ and $c/a$ are about 0.1 larger at the end of the
simulation.  This difference must be due to orbit scattering (either 
physical or purely numerical); the fact that the difference between 
these two runs is so much smaller than that between $t=0$ and $t_g$ implies 
that scattering has only a mild effect on the halo shape.  The global 
shape change at $t_g$ can therefore be attributed to orbit deformation.

Model SA1 had a disk with $R_{\rm d} = 3 \kpc$ and a baryon-to-dark
matter fraction, $f_b = 0.039$, consistent with estimates for local
galaxies \citep{jim_etal_03}.  A more massive or more compact galaxy
may lead to greater scattering.  We explored to what extent larger
$f_b$ or smaller $R_{\rm d}$ affect the halo shape in two further
simulations.
Run SA2 increased $M_b$ by a factor of 3 while keeping $R_{\rm d}$
fixed.  The halo shape is changed significantly all the way out to
$r_{200}$ once the disk is evaporated, but remains quite prolate, with
$b/a < 0.6$ and $c/a < 0.5$, as can be seen in Figure
\ref{fig:shapes}b.  In contrast, at $t_g$ the halo has $0.5 < b/a <
1.0$ within the inner 50 kpc.  Even with this high $f_b \simeq 0.12$,
or $\sim 70\%$ of the full cosmic baryon fraction \citep{spe_etal_06},
the irreversible change to the halo shape is $\la 50\%$ of the full
change at $t_g$ out to 100\kpc.  Run SA3 instead set $R_{\rm d} =
1.5\kpc$, keeping the ratio $z_{\rm d}/R_{\rm d}$ fixed (and
decreasing all softenings appropriately).  The evolution in this case
is shown in Figure \ref{fig:shapes}c; as in run SA1, although the halo
at $t=t_g$ is substantially rounder than at the start, after the disk
is evaporated the halo recovers most of its original shape.  Of runs
SA2 and SA3, making the disk more massive (SA2) produced a larger
irreversible change in the halo than did making it more compact.

Run IA1 explored whether having the disk orthogonal to the
intermediate axis makes a significant difference to the halo
distribution, with all other parameters as in run SA1.  The resulting
shape evolution is presented in Figure \ref{fig:shapes}d.  In this
simulation the halo at $t_g$ remained more elongated than that in run SA1
despite having the same $M_b$.  The axis ratios of the halo at $t_g$
cross over at $\sim30 \kpc$, where the halo's flattening orthogonal to
the disk causes the minor axis to switch from the disk plane to the
orthogonal direction.  Once the disk is evaporated, the halo ends very
nearly axisymmetric in cross section in this inner region but
continues to be highly prolate.  As in run SA1, the net change in halo
shape is relatively small at $t_g + t_e$.

\subsection{Long-Axis Experiments}

In run LA1 we placed the disk with its symmetry axis along the long
axis of the halo.  This orientation has been suggested to be favored
by the distribution of satellites around the Milky Way
\citep{zen_etal_05} and by the Sagittarius dwarf tidal stream
(\citet{helmi_04} but see \citet{fel_etal_06} for a different view).  Other 
than the disk's orientation, the parameters of this model are identical to
those of run SA1.  As in that model, the halo in run LA1 is
significantly deformed to large radius by the growing disk, but it
recovers its shape nearly completely once the disk is evaporated, as
shown in Figure \ref{fig:shapes}e.  Likewise, the spherically averaged
kinematic evolution of run LA1 is indistinguishable from that of SA1,
as seen in Figure \ref{fig:kinematics}.

\begin{figure*}
{
\includegraphics[width=\hsize]{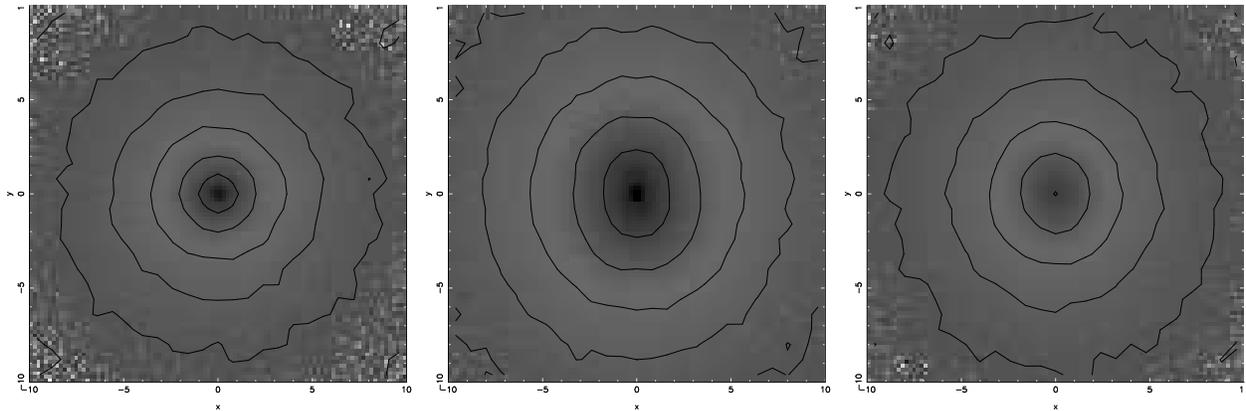}
}
\caption{Evolution of the inner halo in run LB1, seen in cross section
in the disk plane, with only the region $|z| < 5 \kpc$ shown, where
the $z$-axis is the symmetry axis of both disk and halo.  The panels
show $t=0$ ({\it left}), $t_g$ ({\it middle}) and $t_g+t_e$ ({\it right}).  
The halo is initially axisymmetric, becomes elongated orthogonal to 
the symmetry axis at small radii at $t_g$, and largely recovers its 
axisymmetry at $t_g + t_e$.
\label{fig:3ddensity419}}
\end{figure*}

\begin{figure}
{
\includegraphics[width=\hsize]{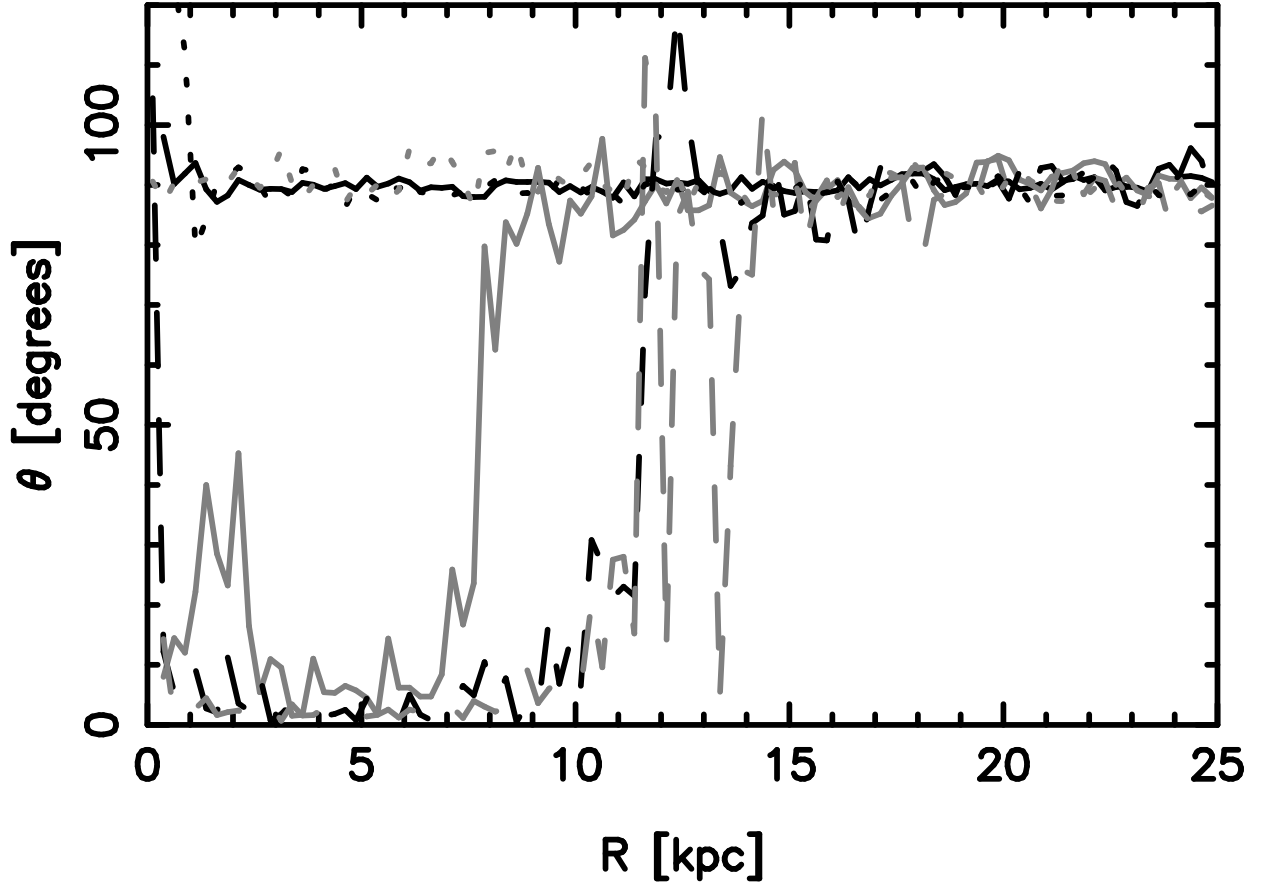}
}
\caption{Major axis orientation evolution in run LB1.  The solid
black and gray lines are at $t=0$ and $t_g/3$, the dashed black and
gray lines are at $2 t_g/3$ and $t_g$ and the black dotted line is at
$t_g+t_e$.  The gray dotted line shows run P$_l$B1 at $t_g$.
\label{fig:angles419}}
\end{figure}

A unique characteristic of the evolution in long-axis experiments is
their tendency for the major axis of the inner halo to switch
orientation by $90\degrees$ into the disk plane once the disk grows
sufficiently massive.  For run LA1 this is evident in Figure
\ref{fig:shapes}e, which shows that the halo is axisymmetric at $\sim
20 \kpc$ (the solid blue line approaches $b/a \simeq 1$, while the
dot-dashed blue line approaches $T \simeq 0$) but is quite
prolate-triaxial at smaller radii.  Major axis flips are more
clearly illustrated by the nearly axisymmetric halo B.  In run LB1, as
\mdisk\ increases, shells of the prolate inner halo become spherically
symmetric.  Further increase in \mdisk\ then leads to the shell
becoming not only flatter vertically (relative to the disk) but also
acquiring an {\it elongation} with its major axis in the plane of the
disk, \ie\ the symmetry axis of the inner halo flips by $90\degrees$
and becomes orthogonal to that of the outer halo (see Figs.
\ref{fig:3ddensity419} and \ref{fig:angles419}).  The direction along
which this reorientation occurs is not random since the halo is
initially not perfectly axisymmetric on large scales.  Continued
increase in \mdisk\ causes the symmetry axis to flip orientation to
larger radii, eventually saturating at $\sim 10$ \kpc.
The halo orientation flips do not occur when the disk is replaced by a
point particle.  The dotted gray line in Figure \ref{fig:angles419}
shows the orientation of the major axis in run P$_l$B1, at a time when
its mass is the same as that in LB1; no flip in the major axis
direction can be seen.

\subsection{Inclined Disk}

\begin{figure}
\centerline{
\includegraphics[width=0.5\hsize]{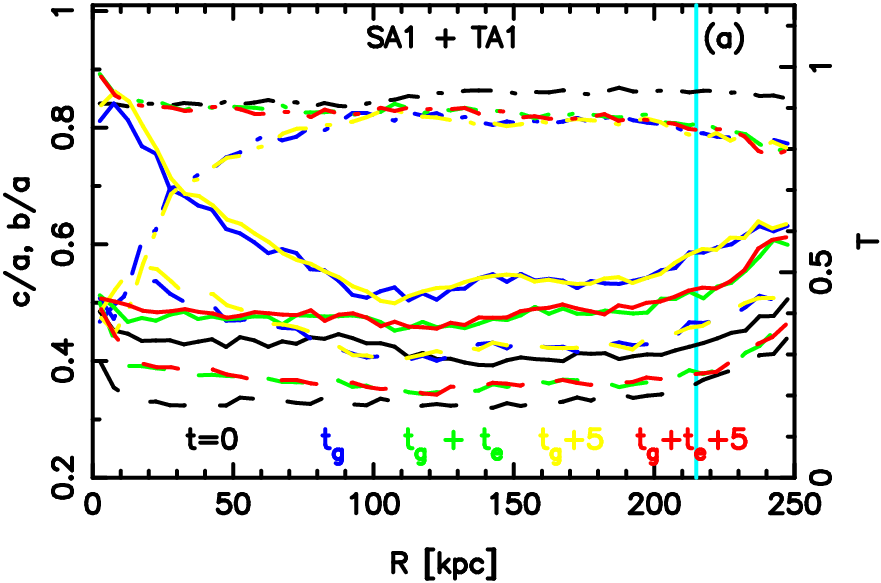}
}
\centerline{
\includegraphics[width=0.5\hsize]{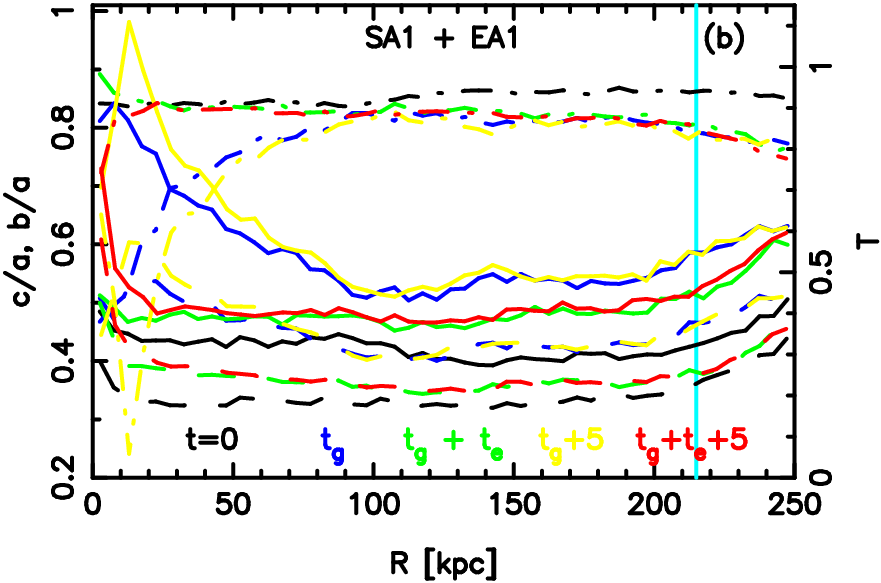}
}
\centerline{
\includegraphics[width=0.5\hsize]{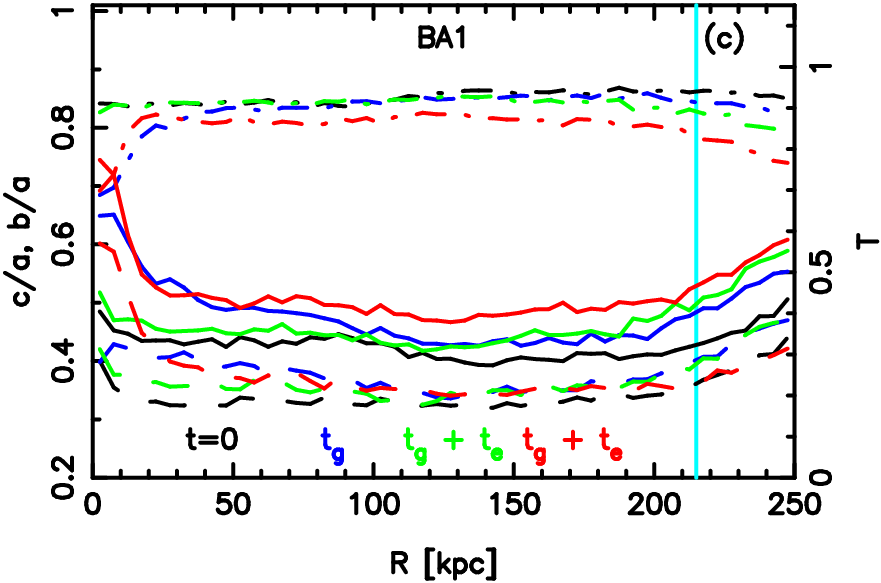}
}
\caption{Shape evolution in runs ({\it a}) TA1, ({\it b}) EA1, and ({\it c}) 
BA1.  The
solid lines show $b/a$, the dashed lines show $c/a$ and the dot-dashed
lines show $T$ (with scale indicated on the right hand side of each
panel).  The black, yellow, and red lines show $t=0$, $t_g$, and $t_g +
t_e$.  For comparison, the equivalent evolution for run SA1 is shown
by the blue and green lines.
\label{fig:nonsymmetric}}
\end{figure}

Simulations have found that the angular momenta of the halo and 
gas need not be aligned \citep{vdb_etal_02, che_etal_03},
although the inner halo and the disk that would form settle to a
common plane \citep{dub_kui_95}.  Therefore we also explored the
effect of a disk inclined relative to the main plane of the dark
matter halo.  Run TA1 is based on run SA1 but with the disk inclined by
$30\degrees$ about the $y-$axis (intermediate axis), all other
parameters being the same.  The resulting evolution is virtually
indistinguishable from that in run SA1 at all times, as shown in
Figure \ref{fig:nonsymmetric}a.

\subsection{Elliptical Disk}

A disk forming in an elliptical potential becomes elongated with its
major axis orthogonal to that of the potential
\citep[e.g.,][]{ger_vie_86}.  In run EA1 we replaced the disk in run
SA1 by an elliptical disk, with its long axis along the $y-$axis.  We
obtain this oval disk by shrinking the $x$ coordinates (parallel to
the halo major axis) of all disk particles by a factor of 0.75;
i.e., the ellipticity of the disk density was $\epsilon = 1 - b/a =
0.25$.  The degree to which
elliptical disks change halo distributions is over estimated by this
simulation since the disk ellipticity is quite large for a massive
disk.
We then evolved this system identically to run SA1, keeping
the disk fixed in place.  Figure \ref{fig:nonsymmetric}b compares the
shape evolution with that in run SA1.  At $t_g + t_e$ the halo is left
significantly rounder within $\sim 20 \kpc$ than it was in run SA1,
but beyond that the evolution is very similar.

\subsection{Central Softened Point Masses}

All of the experiments described above had rigid disks frozen in
place.  While we have been careful to recenter the halo in position
and velocity after the mergers and before growing the disks, some
residual relative motion of the inner and outer parts of the halos
remained.  This motion is damped as the mass of the disk increases,
possibly causing some scattering of orbits.  In order to test for
artifacts associated with such damping, we resorted to simulations
with only a single baryonic particle and compared the evolution when
the particle is free to move (P$_l$B1) and when it is frozen in place
(P$_f$B2).  The rotation curve at $t_g$ is shown in Figure
\ref{fig:rotcurves}.  Figure \ref{fig:shapes}g shows their shape
evolution; in both cases, the shape is largely recovered at $t_g +
t_e$.  If anything, P$_l$B1 is very slightly rounder at 10 \kpc\ than
P$_f$B2 compared with the initial halo.  Figure
\ref{fig:kinematics} shows that their density and kinematic evolution
also is largely reversible.  Thus, our use of rigid disks nailed in
place could not have induced much artificial orbit scattering.

\subsection{Ultrahard particle}

In runs P$_l$B1 and P$_f$B2 the growing particle had a softening of
$\epsilon = 3 \kpc$, a reasonable size for a galaxy.  In run P$_l$B3
we decreased the softening length of the particle to $100 \pc$ and
$M_b$ by half.  Despite the smaller $M_b$ the final halo after $t_g +
t_e$ remains substantially rounder inside $20 \kpc$ than in those runs
(but is largely recovered at larger radii).  If the central particle
were a black hole, its sphere of influence assuming $\sigma_0 = 100
\kms$ from $t=0$ would be $\sim 15 \kpc$.  This is comparable to the
radius out to which the particle irreversibly alters the shape of the
halo.  Likewise, the halo mass within $20\kpc$ is comparable to that
of the central particle: at $t=0$, the halo mass within this radius is
$4 M_b$.
Unlike runs P$_l$B1 and P$_f$B2, Figure \ref{fig:kinematics} shows
that the kinematic evolution is not reversible, and the halo of run
P$_l$B3 becomes significantly tangentially anisotropic, as expected if
box orbits are destroyed.
Despite the different final state, Figure \ref{fig:shapes}h also shows
that the halo shape at $t_g$ is not much different from that in
P$_l$B1 and P$_f$B2, implying that halo shape change is not dominated
by scattering.
Whereas run SA3 with $R_{\rm d} = 1.5 \kpc$, which is not unreasonably
small for most galaxies, did not significantly cause box orbit
destruction, the $\sim 10$ times more centrally concentrated run
P$_l$B3 is able to cause a large irreversible change to the halo shape
out to $\sim 0.3~ r_{200}$.  However, $R_{\rm d}/r_{200} \simeq 0.001$
is unrealistically small.


\section{Angular Momentum Transport}
\label{sec:angmomtran}

We next explore the effect of angular momentum transfer to the halo.
Such transfer is irreversible so the change inflicted on the halo must
also be irreversible.  How strongly the halo distribution is changed
depends on the mechanism by which angular momentum is transferred.  If
via bars or spirals then we may expect that the changes
are mostly at small radius.  If
angular momentum is transferred by satellite galaxies, however, then
the effect on the halo is likely to be much more widespread.

\subsection{Live Barred Disk}

In run BA1 we evolved a model with a live disk for 10 Gyr after $t_g$
before evaporating the disk.  The initial system was similar to run
SA1 but with only $30 \%$ of its $M_b$.  We chose this lower mass
because the same mass as SA1 leads to a long-lived bar, whereas we are
interested in forming a bar that gets destroyed in order to be able
to evaporate the disk.  A bar quickly formed and was subsequently
destroyed \citep{ber_etal_06}.  After 10 Gyr, we fixed the disk
particles in place and evaporated the disk.
Very little of the inner halo shape is recovered after the disk is
evaporated.  In the inner $\sim 20$ kpc the halo remains rounder than
at $t=0$, with both $b/a$ and $c/a$ larger by $\ga 0.1$.  The
irreversible change in the halo is associated with the transfer of
angular momentum from disk to halo \citep{weinbe_85, deb_sel_98}.
Run BA1 produces a comparable change in the inner 20 kpc of the halo
as did the $10$ times more massive run SA2, but the shape change is
much smaller farther out.  Since the bar transfers angular momentum to the
halo at resonances \citep{weinbe_85}, and the strongest of these are at 
smaller radii (most of the angular momentum gained by the halo is 
within the inner $\sim 10 \kpc$), this accounts for the relatively small
radial extent of the halo shape change.

\subsection{Satellites}

Baryons need not cool directly onto the central disk but 
onto satellites instead.  The presence of large numbers of dark
satellites is one of the main predictions of CDM \citep{moo_etal_99,
kly_etal_99, ghi_etal_00}.  As they sink, satellites lose angular
momentum to the halo; in the process box orbits may be scattered.
We explored this evolution with models P$_l$A1 and P$_l$A2.  Starting
with halo A, we selected 10 particles that stay within 200 \kpc\ but
otherwise at random, and adiabatically increased each of their masses
to give the same total baryonic mass as in run SA1.  We grew these 
satellites to full mass and then evaporated them.  Each satellite had
$\epsilon = 0.5 \kpc$ in P$_l$A1 and $5 \kpc$ in P$_l$A2.  Since we
only used softened point particles as satellites, which cannot be
tidally stripped, their effect on the halo is larger than it would be
in nature.  Of the 10 satellites, only one remained at $r > 50 \kpc$,
the rest having fallen to $R < 25 \kpc$ by the end of the simulation.
The evolution of the halo shape in these two models is presented in
Figure \ref{fig:shapes}f.  After the particles reach their full mass,
the halo of run P$_l$A1 is about as round as that in run SA1.  However, the
halo does not recover much of its original shape after the particles
are evaporated.  Clearly, the distribution function of the halo has
been altered to a large extent.
Figure \ref{fig:kinematics} shows that angular momentum transferred by
baryons to the halo can erase the cusp, in agreement with previous
results \citep{ton_etal_06, mas_etal_06, rea_etal_06, wei_kat_07}
although the contraction caused by the growing central mass masks the
core.
The halo shape change at $t_g + t_e$ is very similar in the two runs
because of quite similar angular momentum absorbed by the halo.  
Because of the difference in softenings, the different baryonic potential 
at the halo center at $t_g$ accounts for the $b/a \sim 0.1$
difference in shape, with the softer potential supporting the more
elongated shape.


\section{Orbital Evolution}
\label{sec:orbitanalysis}

We explored directly the evolution of the orbital character of the
models by considering a subsample of 1000 particles in run SA1 and
following their orbits at various points in the simulation.  The 1000
particles were randomly chosen from the $t=0$ distribution such that
they were inside $r=200\kpc$.  We then integrated their motion as test
particles while holding all the other particles fixed in place.  We
used a fixed timestep of 0.1 Myr and integrated for 15 Gyr, storing
the phase space coordinates of each test particle every 1 Myr.  For
the same 1000 particles, we carried out this operation at $t=0$,
$t_g$, and $t_g+t_e$.  Because we froze the background potential, in
effect we have computed the orbital character of the particles at
these three times.  The fact that we integrated for 15 Gyr ensures
that we have sufficient points on each orbit to properly characterize
it.
In this paper we demonstrate with a few examples that a large fraction
of box-like orbits at $t=0$ return to very similar box-like orbits at
$t_g+t_e$, showing that deformation, not transformation, is
responsible for shape change in most cases.  We do this by presenting
their configuration space projection at each of the three different times.
Such an analysis cannot distinguish between box orbits and mildly
chaotic, elongated orbits but this is unimportant anyway for our
present purposes since we have integrated for over a Hubble time.  If
they are mildly chaotic, they can still support a triaxial halo.  A
full analysis of the orbital structure using more sophisticated
techniques will be presented elsewhere.

\begin{figure*}
{
\includegraphics[width=\hsize]{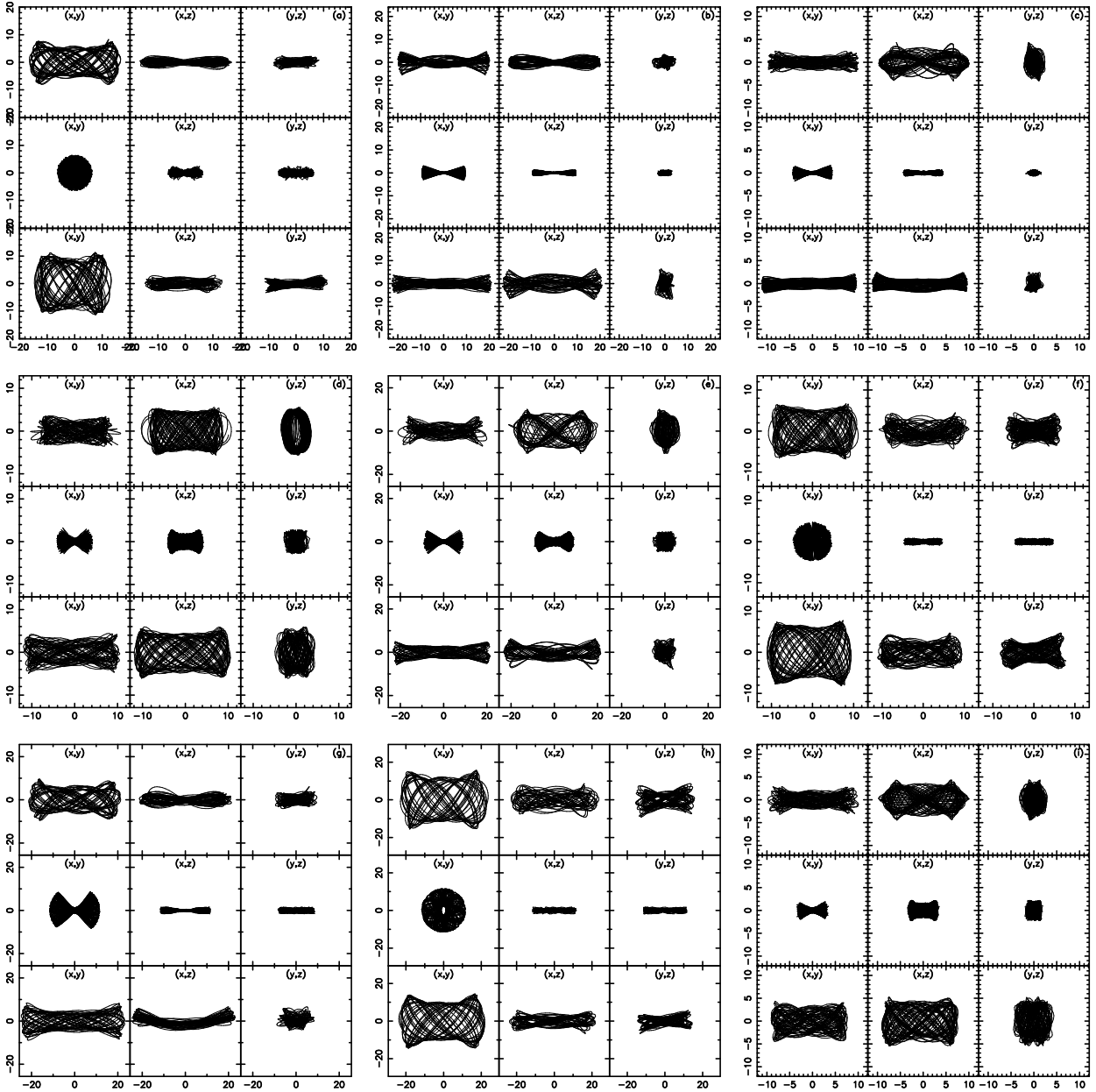}
}
\caption{Sample of initially boxlike orbits in run SA1.  In each
set of panels, the top row is at $t=0$, the middle row is at $t_g$, and
the bottom row is at $t_g+t_e$; from left to right the panels show
projections onto the $(x,y)$, $(x,z)$, and $(y,z)$ planes.  Each orbit
has been integrated for 15 Gyr from each of $t=0$, $t_g$ and
$t_g+t_e$.
\label{fig:orbits563}}
\end{figure*}

Of the 1000 orbits, we start by presenting nine particles, initially on 
boxlike orbits, defined such that, at $t=0$, they
(1) remain inside $25 \kpc$, (2) do not have a fixed sense of rotation
relative to any of the three major axes, (3) reach a radius of at
least $10 \kpc$, and (4) get within $0.2 \kpc$ of the center.
The evolution of many of the other 991 orbits is qualitatively similar
to that of the nine presented here.  Figure \ref{fig:orbits563} projects
these orbits onto the halo symmetry planes, where the $x$-axis is the
halo's major axis and the $z$-axis is the disk's symmetry axis.  Most
orbits at $t_g+t_e$ are quite similar to what they looked like at
$t=0$.  None of the orbits seem strongly chaotic, neither at $t_g$ nor
at $t_g + t_e$, although they may be weakly chaotic.  Moreover, most
orbits retain a box-like shape at $t_g$, but have a significantly
rounder shape than those at $t=0$.  At $t_g$, three of the initially
box-like orbits become round (orbits "a", "f" and "h"); of
these only orbit h changes character completely, becoming a loop
orbit.  Some of the orbits have a slight banana shape; in the full
sample of orbits we found many cases of strongly banana-shaped orbits.
These had a tendency to become more planar but are still distinctly
elongated at $t_g+t_e$.  In a few cases we also found the opposite
occurring --- slightly banana orbits becoming more strongly curved ---
but this was less common.  Of the box-like orbits in Figure
\ref{fig:orbits563} some are rounder in the $(x,y)$ plane at $t_g+t_e$
(\eg, orbits a, f and i), but some are rounder at $t=0$
(\eg, orbits c and e), suggesting that differences in shape are due to
scattering.  What little difference in orbit shape occurs between
$t=0$ and $t_g+t_e$ can probably be attributed to numerical noise.
Most importantly, while there is a clear orbit shape deformation at
$t_g$, little orbital transformation has occurred.

\begin{figure}
{
\includegraphics[angle=-90.,width=\hsize]{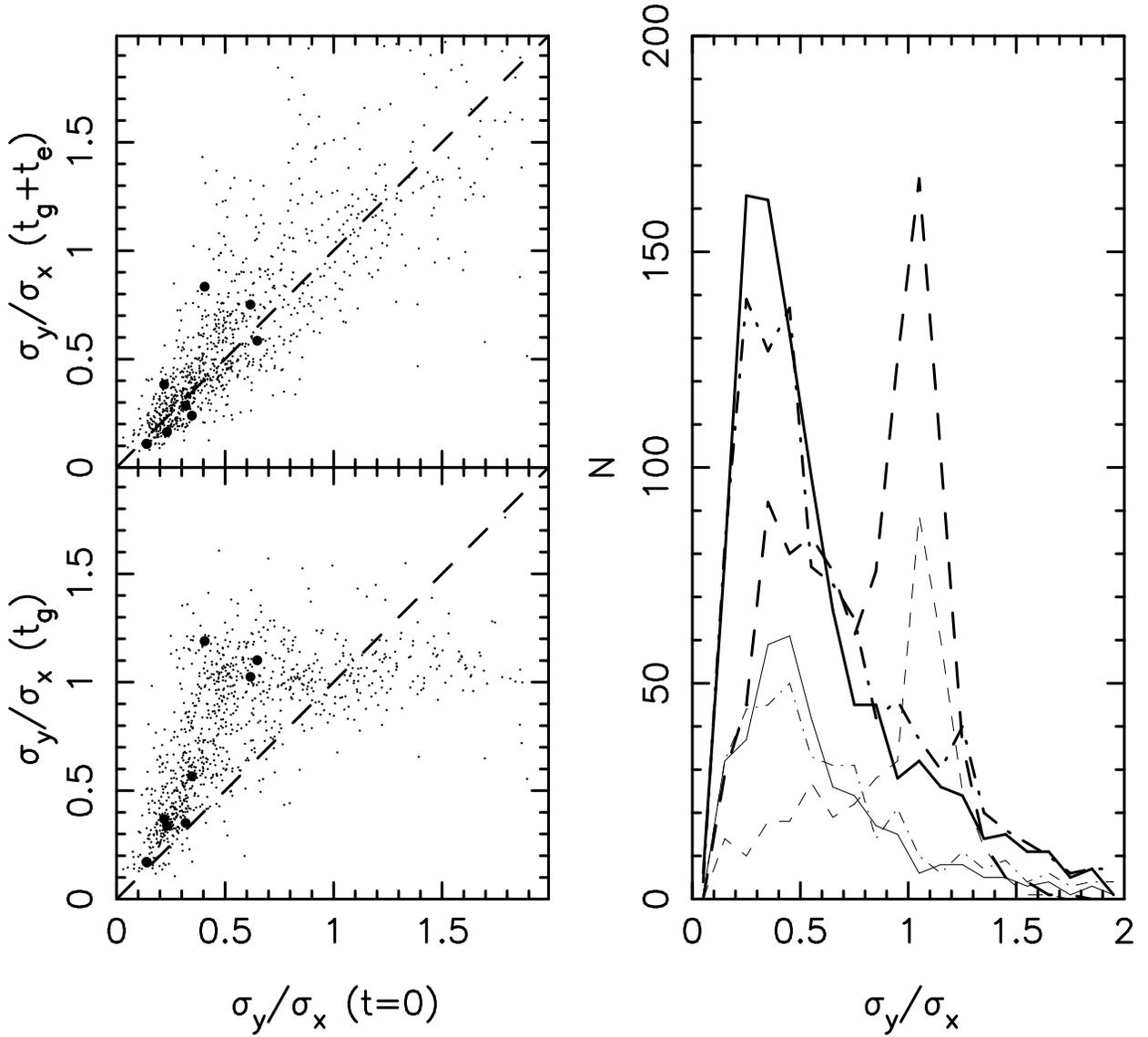}
}
\caption{Deformation of orbits in run SA1.  In the left panels,
the dashed line shows the diagonal.  The solid points show the nine
particles of Fig. \ref{fig:orbits563}.  In the right panel, the
solid, dashed, and dot-dashed lines show the distributions at $t=0$,
$t_g$, and $t_g+t_e$, respectively.  The thick lines are for the full
distribution of 1000 particles, while the thin lines are for those
particles that remain inside $r=50\kpc$ during the 15 Gyr integration
starting at $t=0$.
\label{fig:orbittrans563}}
\end{figure}

We quantify the orbital deformation of the sample of 1000 particles by
plotting in Figure \ref{fig:orbittrans563} $\sigma_y/\sigma_x$,
where $\sigma_x^2 = \sum_t x_t^2$ and similarly for $\sigma_y^2$ and
the sum is over timesteps.  The significantly rounder shape of orbits
at $t_g$ than $t=0$ is apparent, with the vast majority of
orbits initially aligned with the halo having $\sigma_y/\sigma_x$
closer to unity at the later time.  Orbits initially elongated along
the halo's minor axis, as well as orbits initially rounder than
$\sigma_y/\sigma_x \ga 0.6$ end up round, with $\sigma_y/\sigma_x
\simeq 1$.  Instead at $t_g + t_e$ the orbits tend to return to their
initial elongation, especially for the most elongated orbits.  The
right panel shows the distribution of $\sigma_y/\sigma_x$; orbits
become substantially rounder at $t_g$ but the population as a whole
recovers the original distribution to a large extent once the disk is
evaporated.

\begin{figure*}
{
\includegraphics[width=0.5\hsize]{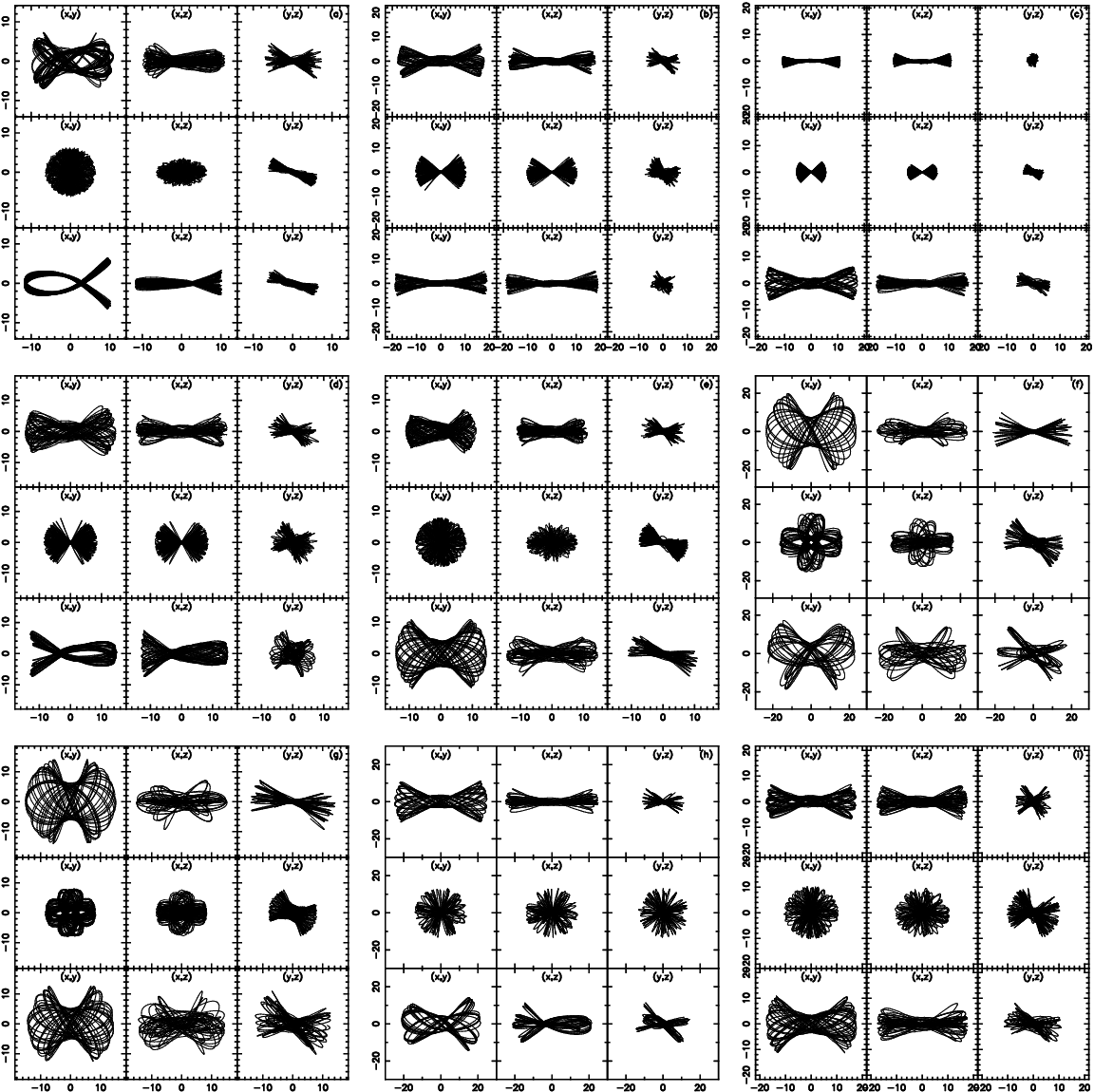}
\includegraphics[width=0.5\hsize]{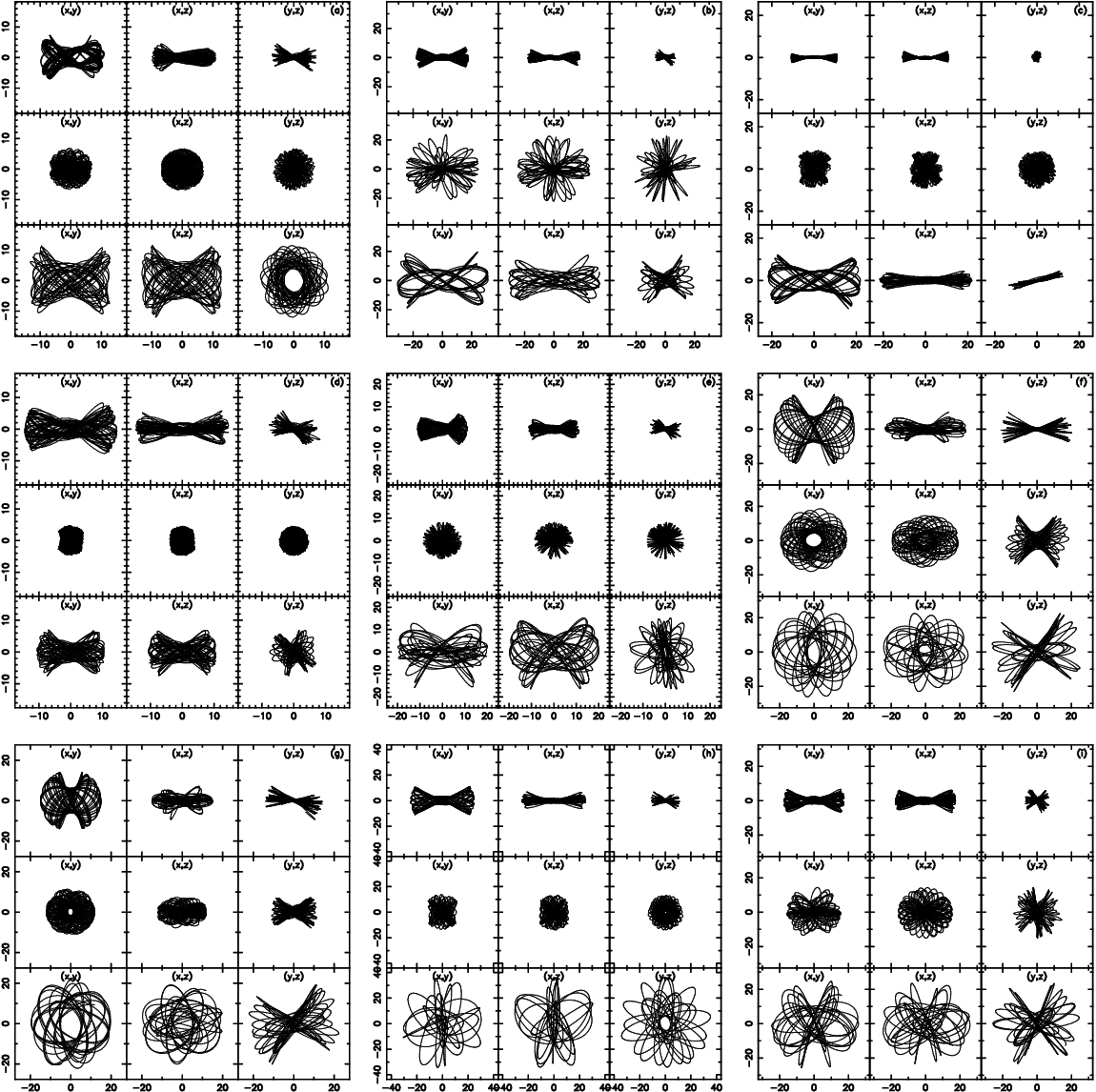}
}
\caption{Sample of initially boxlike orbits in run P$_f$B2 ({\it left three
columns}) and P$_l$B3 ({\it right three columns}).  We show the evolution of the
same nine starting orbits in the two different models.  In each set of
panels, the top row is at $t=0$, the middle row is at $t_g/2$
(P$_f$B2) or $t_g$ (P$_l$B3), and the bottom row is at $t_g+t_e$; from
left to right the panels show projections onto the $(x,y)$, $(x,z)$,
and $(y,z)$ planes.  Each orbit has been integrated for 15 Gyr from
each starting point.
\label{fig:orbits448}}
\end{figure*}

\begin{figure*}
{
\includegraphics[angle=-90.,width=0.5\hsize]{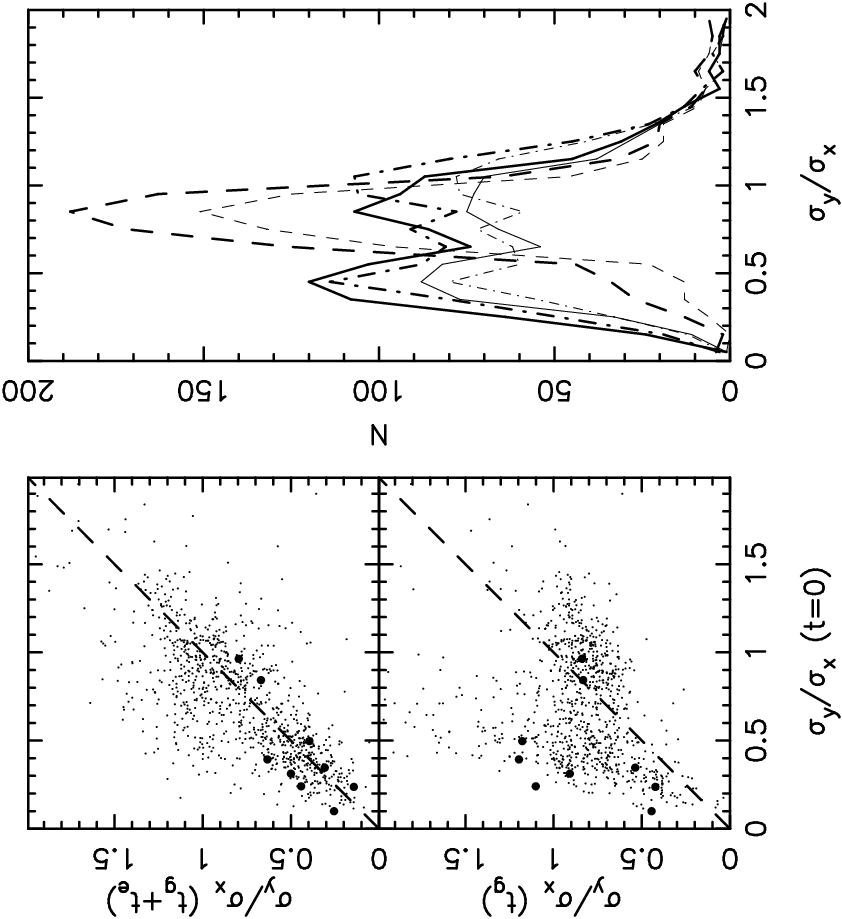}
\includegraphics[angle=-90.,width=0.5\hsize]{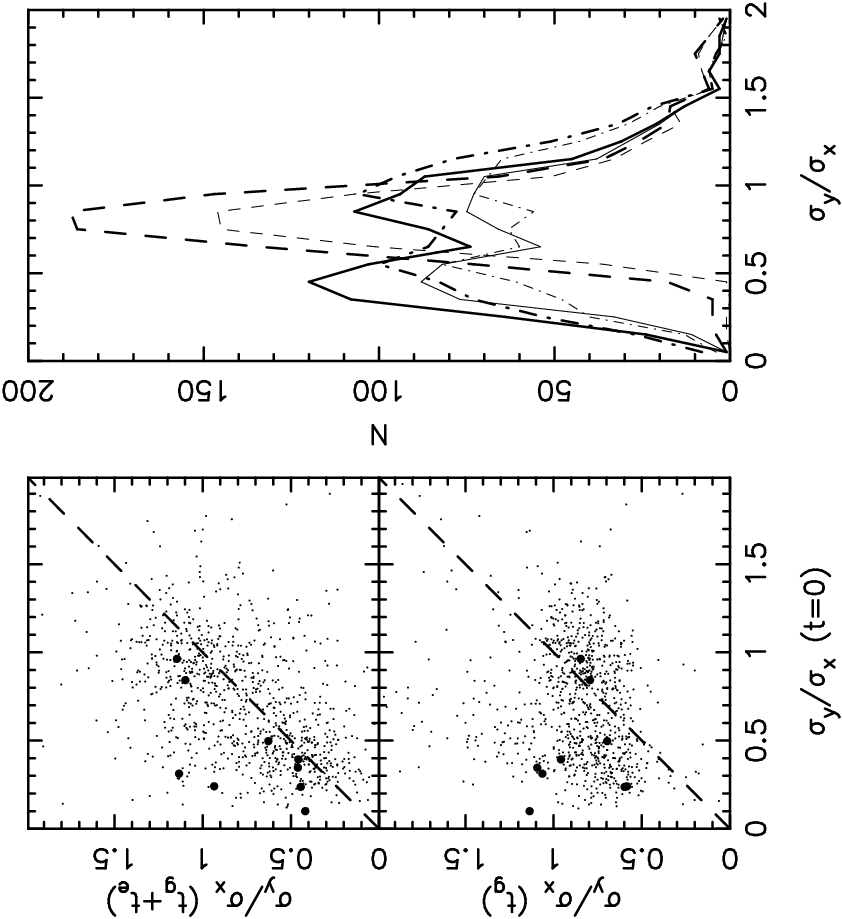}
}
\caption{Identical to Fig. \ref{fig:orbittrans563} but for runs
P$_f$B2 ({\it left}) and P$_l$B3 ({\it right}).  The final distribution 
of orbits
is significantly depleted of elongated orbits in run P$_l$B3 compared
with run P$_f$B2.  All nine orbits of Fig. \ref{fig:orbits448} are
above the diagonal in run P$_l$B3 at $t_g+t_e$ but scatter about the
diagonal in run P$_f$B2.
\label{fig:orbittrans448}}
\end{figure*}

A comparison of orbital evolution in runs P$_f$B2 and P$_l$B3 provides
an example of orbit transformation.  We again selected 1000 orbits
from particles within the inner $200 \kpc$ of halo B at $t=0$.  We
integrated their orbits as above but we used the smaller timestep
$\delta t = 10^4 $ years in the case of P$_l$B3; for P$_f$B2 we use
$t_g/2$ when the central particle has the same mass as at $t_g$ in
model P$_l$B3.  Although we are comparing the two models at the same
central particle mass, the orbits at time $t_g + t_e$ in model P$_f$B2
were computed after the central particle was evaporated from a mass
twice that reached in P$_l$B3.  As before, we present in Figure
\ref{fig:orbits448} nine orbits that at $t=0$ are box-like.  
These boxlike orbits in P$_l$B3 are more often transformed
than those in run P$_f$B2.
In model P$_f$B2, only one orbit (orbit f) appears to have
changed substantially at the end of the simulation, while orbits a
and d become fish orbits (although they may have been
librating about fish orbits at $t=0$).  Orbit b is largely unchanged and the
remaining orbits are all boxlike.  In model P$_l$B3, orbit e is
changed about as much as orbit f in P$_f$B2.  However, four of the nine
orbits, f-i, are very strongly transformed by $t_g + t_e$ and are
no longer able to support a triaxial shape.
Figure \ref{fig:orbittrans448} compares the distribution of all 1000
particles and a depletion of elongated orbits is evident in P$_l$B3
compared with run P$_f$B2.

\section{Discussion}
\label{sec:conc}

\subsection{Timesteps}

\begin{figure}
{
\includegraphics[width=\hsize]{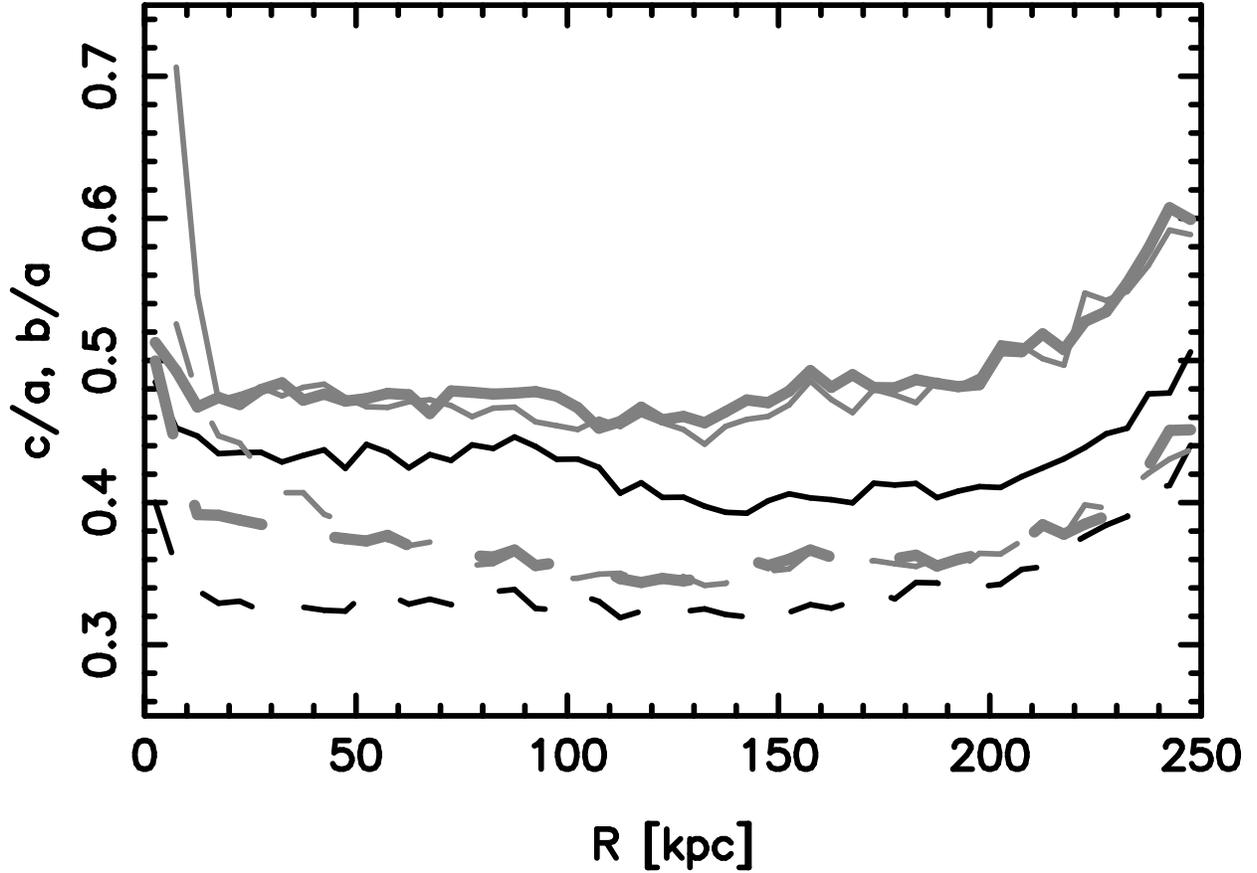}
}
\caption{Effect of timestep size on the evolution of run SA1.  The
black lines shows the shape at $t=0$ while the gray lines are for $t_g
+ t_e$, with the thick line the standard result with $\eta = 0.2$ and
the thin lines for $\eta = 2$.  As in Fig. \ref{fig:shapes}, the
solid lines show $b/a$ while the dashed lines show $c/a$.  The larger
timesteps lead to an evolution that is less reversible.
\label{fig:tstest}}
\end{figure}

We also performed a number of
tests of the numerics to verify that our results are robust.  The main
concern is the timestep used.  \citet{she_sel_04} found that too large
timesteps result in bars being destroyed too easily, because orbits
are not followed accurately near the central mass.  Our simulations
used multi-stepping.  With a base timestep $\Delta t$, particles move
on timesteps $\Delta t/2^n$, where $n$ is the rung level satisfying
the condition $\delta t = \Delta t/2^n < \eta (\epsilon/a)^{1/2}$, and
$\epsilon$ is the particle's softening, $a$ is its acceleration and
$\eta$ a tolerance parameter.  We used $\eta = 0.2$, a conservative
value; with a base timestep $\Delta t = 5$ Myr, simulation SA1 at
$t_g$ had a range of timesteps down to $5/2^5 = 0.16$ Myr.  If instead
we set $\eta = 2$ the timestep distribution only reaches to $5/2 =
2.5$ Myr.  The effect of these larger timesteps, shown in Figure
\ref{fig:tstest}, is manifest at $r \la 20 \kpc$, which remains
significantly rounder at $t_g+t_e$ than when $\eta = 0.2$.  The quite
modest net shape change in our simulations implies that the timesteps
we used were sufficiently small to correctly follow the evolution near
the center.

\subsection{Evidence against box orbit destruction}

We have demonstrated that the substantial axisymmetrization caused by
a disk growing inside a dark matter halo is largely, although not
wholly, reversible.  If triaxial halos become rounder because of box
orbit destruction \citep[\eg][]{mac_etal_07}, then these orbits would
have to be repopulated in order for the halo to recover its original
shape once the disk is evaporated.  Apart from being unlikely for such
a highly ordered system as a triaxial halo, this interpretation is not
supported by our orbital analysis, which shows that box-like orbits
are deformed by realistic disks but not transformed by scattering into
new orbits.  The strongest evidence that deformation is a more
important process than transformation comes from the ability of
particles to return, after the disk is evaporated, to nearly the same 
orbits as they started from.  This makes it implausible that the
shape change is due to a large increase in strong chaos.  Instead we
find that realistic axisymmetric disks are not concentrated or massive
enough to cause substantial chaos.  In a similar vein,
\citet{hol_etal_02} found that chaos in triaxial ellipticals is
induced by the black hole only, not by the stellar cusp surrounding
it.  This need not mean, however, that chaos is not enhanced by the
disk.  It could well be that orbits are becoming
weakly chaotic but do not diffuse sufficiently on a Hubble timescale
to significantly weaken triaxiality.  The shape
evolution due to baryons cooling onto a central galaxy can therefore 
be computed from adiabatic invariants.

The primary role of orbit deformation over transformation is also
indicated by the much smaller effects of scattering when it is clear
that scattering has occurred.  In run SA1 scattering occurred when the
disk was maintained at full mass for an additional 5 Gyr before being
evaporated.  This only caused a small additional change in the final
halo shape.  Furthermore, comparing runs P$_l$B1 and P$_l$B3, we find
that the halo shape at $t_g$ is rather similar, despite the fact that
in run P$_l$B3 box orbits are significantly destroyed.  The same is
true for runs SA1 and P$_l$A1.  These examples directly illustrate
that box orbit destruction is a much smaller factor in halo shape
changes than is orbit deformation.

Weak chaos is consistent with the orbital characterization of
\citet{mac_etal_07} but does not support their claim that the shape
change is largely due to enhanced chaos from the central baryonic
mass.  Another possibility is that in their simulations chaos could
have been caused by gas cooling inside subhalos, which we showed leads
to a substantial change in the phase space distribution of the halo.

\subsection{Implications for galaxy formation simulations}

If baryons cool onto substructures within the halo then box orbits are
very efficiently destroyed.  Since the evolution of disks can be
strongly influenced by halo triaxiality \citep[\eg][]{ide_hoz_00}, any
process that artificially reduces triaxiality can lead to biases in
the properties of galaxies forming in cosmological simulations.
\citet{age_etal_07} demonstrate that the evolution of gas blobs in
smoothed particle hydrodynamics (SPH) simulations is different from
that found in Eulerian gas codes.  They interpreted this difference as
being due to the unphysically poor mixing of traditional SPH, which
allows blobs to survive longer in SPH.  Moreover, satellites in
cosmological simulations tend to have denser, more concentrated gas
components than their real counterparts, which makes them harder to
strip by ram pressure and tides \citep{may_etal_07} and thus more
likely to artificially enhance halo shape changes.  Baryonic cooling
inside substructures is strongly suppressed by feedback from
supernovae \citep{dek_sil_86, gov_etal_04, gov_etal_07}.  If
simulations do not treat feedback properly or have low resolution,
then baryons may condense into concentrated substructures biasing the
global evolution of simulated galaxies.

\subsection{Summary}

Our results can be summarized as follows:

\begin{enumerate}

\item The adiabatic growth of disks with realistic sizes and masses
inside prolate/triaxial halos leads to a large change in the shape of
the halo.  Axis ratios change by $>0.2$ out to roughly $0.5~ r_{200}$.
The growth of the disk drives the halo kinematics to larger radial
anisotropy.  The midplane global potential ellipticity is less than
0.1, consistent with the small scatter in the Tully-Fisher relation
\citep{fra_dez_92}.

\item Despite these large changes, the underlying phase space
distribution is not grossly altered, as we verified by artificially
evaporating the disk and recovering, to a large extent, the original
halo.  The irreversible change in final halo structure is larger for
more massive or more centrally concentrated disks, but is still a
relatively small fraction of the total shape change when the disk is
at full mass.
As in the case of black holes at the centers of cuspy elliptical
galaxies \citep{ger_bin_85, hol_etal_02}, the bulk of the irreversible
halo shape change occurs in the inner region of the galaxy.  This
small irreversible shape change is driven by orbit scattering.

\item Box orbit destruction cannot be the right interpretation for the
shape change caused by disk growth.  Such a process is not reversible
but we found that a large fraction of particles on box-like orbits
individually return to very similar orbits after the disk is
evaporated.  At most only mild chaos is induced.  Instead we find that
box orbits become deformed by the growing disk, but retain their
character, and this seems sufficient to explain the change in shape.
In the absence of angular momentum transport or extreme
mass/concentration galaxies, very little of the quite large shape
change that dark matter halos undergo as baryons condense inside them
is due to box orbit destruction.  As a result, shape change can be well
approximated by adiabatic invariants.

\item Very concentrated structures do lead to scattering and
to large irreversible changes in halo shape and kinematics; we 
found that scales of $\sim 100 \pc$ are needed to
accomplish this.  The irreversible shape change was then
restricted to the sphere of influence of this pointlike mass: 
$r \sim GM/\sigma^2$.  Massive disks are
also able to change the halo structure irreversibly, but the mass
required, $\sim 70\%$ of the cosmic baryon fraction, is quite high.  Even
in such cases of scattering, the halo shape with the baryons at full
mass is not much different from similar simulations with little
scattering, suggesting that it is orbital deformation in the first 
place that changes the shape of the halo.

\item If baryons transport angular momentum to the halo, a large
irreversible change in halo shape and kinematics occurs.  Such
transfers can occur either because of gas condensing in satellites or
nonaxisymmetric structures forming in the disk.  Even quite low mass
disks are able to alter the inner halo distribution if they can
transport angular momentum to the halo.
The effect of satellites can be artificially large in simulations of
galaxy formation because of the poor mixing in SPH \citep{age_etal_07}
and the too highly concentrated satellites that form
\citep{may_etal_07}.

\item When the disk minor axis and halo major axis are aligned, growth
of the disk leads to an elongation within the plane of
the disk, even when the initial halo is very nearly axisymmetric.

\end{enumerate}

\acknowledgements This paper is based in part on work supported by
the National Science Foundation under the following NSF programs:
Partnerships for Advanced Computational Infrastructure, Distributed
Terascale Facility (DTF), and Terascale Extensions: Enhancements to the
Extensible Terascale Facility.  Many additional simulations were
carried out at the University of Z\"urich on zBox and at the Arctic
Region Supercomputing Center.
V. P. D. thanks the University of Z\"urich for hospitality during part
of this project.  V. P. D. is supported by a Brooks Prize Fellowship in
Astrophysics at the University of Washington and receives partial
support from NSF ITR grant PHY-0205413.  S. K. acknowledges support by
the US Department of Energy through a KIPAC Fellowship at Stanford
University and the Stanford Linear Accelerator Center.
We thank Andrea Macci\`{o} and Ioannis Sideris for fruitful
discussions and the anonymous referee for comments that helped to
improve this paper.  V. P. D. thanks Monica Valluri for discussions that
helped improve this paper.


\bibliographystyle{aj.bst}
\bibliography{ms.bbl}
 

\end{document}